\documentclass[longauth]{aa}

\usepackage{graphicx}
\usepackage{txfonts}
\usepackage{natbib}     
\usepackage{xcolor} 
\usepackage[para]{threeparttable} 
\usepackage{float}
\usepackage{threeparttable} 
\usepackage{placeins} 
\usepackage{multirow} 
\usepackage{tablefootnote}

\usepackage{caption}
\usepackage{subcaption}

\usepackage{natbib,twoopt} \usepackage[breaklinks=true]{hyperref}

\bibpunct{(}{)}{;}{a}{}{,} 

\begin{document}

\title{Retrieving day- and nightside atmospheric properties of the ultra-hot Jupiter TOI-2109b}
\subtitle{Detection of Fe and CO emission lines and evidence for inefficient heat transport}

\author{
    D.~Cont \inst{\ref{instLMU}, \ref{instExzO}}
    \and L.~Nortmann \inst{\ref{instIAG}}
    \and F.~Lesjak \inst{\ref{instIAG}}
    \and F.~Yan \inst{\ref{instUSTC}}
    \and D.~Shulyak \inst{\ref{instIAA}}
    \and A.~Lavail \inst{\ref{instIRAP}}
    \and M.~Stangret \inst{\ref{instINAF}}
    \and E.~Pall\'e \inst{\ref{instIAC}, \ref{instULL}}
    \and P.~J.~Amado \inst{\ref{instIAA}}
    \and J.~A.~Caballero \inst{\ref{instCAB}}
    \and A.~Hatzes \inst{\ref{instTLS}}
    \and Th.~Henning \inst{\ref{instMPA}}
    \and N.~Piskunov \inst{\ref{instUppsala}}
    \and A.~Quirrenbach \inst{\ref{instLSW}}
    \and A.~Reiners \inst{\ref{instIAG}}
    \and I.~Ribas \inst{\ref{instICE}, \ref{instIEEC}}
    \and J.~F.~Ag\"u\'i~Fern\'andez \inst{\ref{instCAHA}}
    \and C.~Ak{\i}n \inst{\ref{instLMU}, \ref{instSpaceHabi}}
    \and L.~Boldt-Christmas \inst{\ref{instUppsala}}
    \and P.~Chaturvedi \inst{\ref{instMumbai}, \ref{instTLS}}
    \and S.~Czesla \inst{\ref{instTLS}}
    \and A.~Hahlin \inst{\ref{instUppsala}}
    \and K.~Heng \inst{\ref{instLMU}, \ref{instARTORG}, \ref{instUCL}, \ref{instAAG}}
    \and O.~Kochukhov \inst{\ref{instUppsala}}
    \and T.~Marquart \inst{\ref{instUppsala}}
    \and K.~Molaverdikhani \inst{\ref{instLMU}, \ref{instExzO}}
    \and D.~Montes \inst{\ref{instUCM}}
    \and G.~Morello \inst{\ref{instIAA}, \ref{instINAFPA}}
    \and E.~Nagel \inst{\ref{instIAG}}
    \and J.~Orell-Miquel \inst{\ref{instAustin}, \ref{instULL}}
    \and A.~D.~Rains \inst{\ref{instUppsala}}
    \and M.~Rengel \inst{\ref{instMPS}}
    \and A.~Schweitzer \inst{\ref{instHAM}}
    \and A.~S\'anchez-L\'opez \inst{\ref{instIAA}}
    \and U.~Seemann \inst{\ref{instESO}}
}

\institute{
    Universit\"ats-Sternwarte, Ludwig-Maximilians-Universit\"at M\"unchen, Scheinerstrasse 1, 81679 M\"unchen, Germany\label{instLMU}\\ 
    \email{david.cont@lmu.de}
    \and
    Exzellenzcluster Origins, Boltzmannstrasse 2, 85748 Garching bei M\"unchen, Germany\label{instExzO}
    \and
    Institut f\"ur Astrophysik und Geophysik, Georg-August-Universit\"at G\"ottingen, Friedrich-Hund-Platz 1, 37077 G\"ottingen, Germany\label{instIAG} 
    \and
    Department of Astronomy, University of Science and Technology of China, Hefei 230026, People’s Republic of China\label{instUSTC}
    \and 
    Instituto de Astrof{\'i}sica de Andaluc{\'i}a (IAA-CSIC), Glorieta de la Astronom{\'i}a s/n, 18008 Granada, Spain\label{instIAA}
    \and
    Namzitu astro, F-31130 Quint-Fonsegrives, France\label{instIRAP} 
    \and
    INAF -- Osservatorio Astronomico di Padova, Vicolo dell’Osservatorio 5, 35122, Padova, Italy\label{instINAF}
    \and
    Instituto de Astrof{\'i}sica de Canarias (IAC), Calle V{\'i}a Lactea s/n, 38200 La Laguna, Tenerife, Spain\label{instIAC}
    \and
    Departamento de Astrof{\'i}sica, Universidad de La Laguna, 38026  La Laguna, Tenerife, Spain\label{instULL}
    \and
    Centro de Astrobiolog{\'i}a (CAB), CSIC-INTA, Camino bajo del castillo s/n, Campus ESAC, 28692 Villanueva de la Ca{\~n}ada, Madrid, Spain\label{instCAB}
    \and
    Th\"uringer Landessternwarte Tautenburg, Sternwarte 5, 07778 Tautenburg, Germany\label{instTLS}
    \and
    Max-Planck-Institut für Astronomie, K\"onigstuhl 17, 69117 Heidelberg, Germany\label{instMPA}
    \and
    Department of Physics and Astronomy, Uppsala University, Box 516, 75120 Uppsala, Sweden\label{instUppsala}   
    \and
    Landessternwarte, Zentrum f\"ur Astronomie der Universit\"at Heidelberg, K\"onigstuhl 12, 69117 Heidelberg, Germany\label{instLSW}     
    \and
    Institut de Ci\`encies de l'Espai (CSIC-IEEC), Campus UAB, c/ de Can Magrans s/n, 08193 Bellaterra, Barcelona, Spain\label{instICE}
    \and
    Institut d'Estudis Espacials de Catalunya (IEEC), 08034 Barcelona, Spain\label{instIEEC}
    \and
    Centro Astron{\'o}mico Hispano en Andaluc{\'i}a, Observatorio de Calar Alto, Sierra de los Filabres, 04550 G{\'e}rgal, Spain\label{instCAHA} 
    \and
    Center for Space and Habitability, University of Bern, Gesellschaftsstrasse 6, CH-3012 Bern, Switzerland\label{instSpaceHabi}
    \and
    Department of Astronomy and Astrophysics, Tata Institute of Fundamental Research, Mumbai, India, 400005\label{instMumbai} 
    \and    
    ARTORG Center for Biomedical Engineering Research, University of Bern, Murtenstrasse 50, CH-3008, Bern, Switzerland\label{instARTORG}
    \and
    University College London, Department of Physics \& Astronomy, Gower St, London, WC1E 6BT, United Kingdom\label{instUCL}
    \and
    Astronomy \& Astrophysics Group, Department of Physics, University of Warwick, Coventry CV4 7AL, United Kingdom\label{instAAG}
    \and
    Departamento de F\'{i}sica de la Tierra y Astrof\'{i}sica and IPARCOS-UCM (Instituto de F\'{i}sica de Part\'{i}culas y del Cosmos de la UCM), Facultad de Ciencias F\'{i}sicas, Universidad Complutense de Madrid, 28040, Madrid, Spain\label{instUCM}
    \and
    INAF, Osservatorio Astronomico di Palermo, Piazza del Parlamento 1, 90134 Palermo, Italy\label{instINAFPA}
    \and
    Department of Astronomy, University of Texas at Austin, 2515 Speedway, Austin, TX 78712, USA\label{instAustin}
    \and
    Max-Planck-Institut für Sonnensystemforschung, Justus-von-Liebig-Weg 3, 37077 G\"ottingen, Germany\label{instMPS}
    \and
    Hamburger Sternwarte, Gojenbergsweg 112, 21029 Hamburg, Germany\inst{\label{instHAM}}
    \and
    European Southern Observatory, Karl-Schwarzschild-Str. 2, 85748 Garching bei M\"unchen, Germany\label{instESO}
}

\date{Received 17 March 2025 / Accepted 17 April 2025}


\abstract
{The ultra-hot Jupiter (UHJ) TOI-2109b marks the lower edge of the equilibrium temperature gap between 3500\,K and 4500\,K, an unexplored thermal regime that separates KELT-9b, the hottest planet yet discovered, from all other currently known gas giants. To study the thermochemical structure of TOI-2109b's atmosphere, we obtained high-resolution emission spectra of both the planetary day- and nightsides with CAHA/CARMENES and VLT/CRIRES$^+$. By applying the cross-correlation technique to the high-resolution spectra, we identified the emission signatures of \ion{Fe}{i} (S/N\,=\,4.3) and CO (S/N\,=\,6.3), as well as a thermal inversion layer in the dayside atmosphere; no significant H$_2$O signal was detected from the dayside. None of the analyzed species were detectable from the nightside atmosphere. We applied a Bayesian retrieval framework that combines high-resolution spectroscopy with photometric measurements to constrain the dayside atmospheric parameters and derive upper limits for the nightside hemisphere. The dayside thermal inversion extends from approximately 3200\,K to 4600\,K, with an atmospheric metallicity consistent with that of the host star (0.36\,dex). Only weak constraints could be placed on the C/O ratio, with a lower limit of 0.15. The retrieved spectral line broadening is consistent with tidally locked rotation, indicating the absence of strong dynamical processes in the atmosphere. An upper temperature limit of approximately 2400\,K and a maximum atmospheric temperature gradient of about 700\,K/$\log{\mathrm{bar}}$ could be derived for the planetary nightside. Comparison of the retrieved dayside temperature-pressure profile with theoretical models, the absence of strong atmospheric dynamics, and significant differences in the thermal constraints between the day- and nightside hemispheres suggest a limited heat transport efficiency across the planetary atmosphere. Overall, our results place TOI-2109b in a transitional regime between the UHJs below the thermal gap, which show both CO and H$_2$O emission lines, and KELT-9b, where molecular features are largely absent.}


\maketitle

%

\section{Introduction}

Ultra-hot Jupiters (UHJs) are gas giant exoplanets on close-in orbits around their host stars with dayside temperatures exceeding 2200\,K \citep[e.g.,][]{Parmentier2018, Kitzmann2018}. Their large radii and high atmospheric temperatures result in extended scale heights and strong thermal fluxes, making them ideal targets for atmospheric characterization via transmission and emission spectroscopy. The extreme thermal conditions of UHJs lead to dayside atmospheres dominated by atomic and ionic species \citep[e.g.,][]{BellCowan2018, Arcangeli2018}. Only a few molecular species with strong chemical bonds, such as CO, can survive in significant concentrations under these extreme conditions \citep[e.g.,][]{Parmentier2018, Kreidberg2018}. Observations have identified \ion{Fe}{i} and CO as dominant contributors to UHJ spectra at visible and near-infrared wavelengths, respectively \citep[e.g.,][]{Pino2020, Kasper2021, Yan2022b, Ramkumar2023, Lesjak2025}. Another key characteristic of UHJs is the presence of thermal inversions in their dayside atmospheres, driven by strong absorption of stellar irradiation at visible and ultraviolet wavelengths in the upper atmospheric layers \citep[e.g.,][]{Hubeny2003, Fortney2008, Lothringer2018, Arcangeli2018, Fossati2021, Fossati2023}.

Ultra-hot Jupiters typically exhibit synchronous rotation caused by tidal circularization during the early evolutionary stages of these planetary systems \citep{Hut1981}. Therefore, UHJs have permanent day- and nightsides with significant differences in their thermal properties. The thermal contrast between the two planetary hemispheres depends on the atmospheric heat redistribution efficiency \citep{CowanAgol2011} controlled by global gas flows, which can be classified into two dominant dynamical regimes: day- to nightside winds and super-rotating jet streams, both capable of reaching speeds of several km\,s$^{-1}$ \citep[e.g.,][]{MillerRicciKempton2012, Showman2013, TanKomacek2019, Seidel2025}. With increasing temperature, UHJs experience a decreasing heat transport across the atmosphere as frictional drag from the coupling between the partially ionized atmosphere and the planetary magnetic field, as well as radiative cooling, can efficiently dampen large-scale atmospheric gas flows \citep[e.g.,][]{KomacekShowman2016, Schwartz2017, Arcangeli2018}. The heat transport efficiency reaches a minimum at a dayside-redistribution equilibrium temperature\footnote{As in \cite{Wong2021}, $T_\mathrm{eq}$ refers to the dayside-redistribution equilibrium temperature, which assumes uniform heat redistribution over the dayside and $T$\,=\,0\,K in the nightside hemisphere \citep{CowanAgol2011}.} of $T_\mathrm{eq}$\,$\sim$\,3000\,K, beyond which the atmospheric heat redistribution becomes more effective again \citep[e.g.,][]{Fortney2021, Wong2021}. This is caused by the dissociation of molecular hydrogen into atomic hydrogen at $T_\mathrm{eq}$\,$>$\,3000\,K, which is transported to the nightside by atmospheric winds, where it recombines and releases significant amounts of energy \citep[e.g.,][]{BellCowan2018, KomacekTan2018, TanKomacek2019, Roth2021}.

So far, high-resolution spectroscopy has allowed the successful characterization of the atmospheric conditions in the daysides and at the terminators of UHJs \citep[e.g.,][]{YanHenning2018, Pino2020, Kasper2021, Prinoth2022, Bello-Arufe2022a, Zhang2022, Brogi2023, Johnson2023, Pelletier2023, Borsato2023a, Hoeijmakers2024, Ramkumar2025}. However, high-resolution spectroscopy studies focusing on the physical and chemical conditions in the nightside atmospheres of exoplanets remain limited. In the literature, only tentative evidence for nightside emission has been reported for two exoplanets, the hot Jupiter HD\,179949b \citep{Matthews2024} and the UHJ WASP-33b \citep{Yang2024, Mraz2024}, with the latter showing conflicting results between the two existing studies. In contrast, space-based photometric and low-resolution spectroscopy measurements have successfully detected the thermal flux and placed constraints on the brightness temperature values for exoplanet nightside atmospheres \citep[e.g.,][]{Keating2019, Morello2019, Morello2023}. The difficulty of making similar detections with ground-based high-resolution spectroscopy likely stems from the absence of strong line features in the emission signature of UHJ nightsides, as the technique is only sensitive to spectral lines but insensitive to broadband features in the spectra. In fact, theoretical models predict a significant flattening of the temperature-pressure ($T$-$p$) profiles towards the upper atmospheres of UHJ nightsides \citep[e.g.,][]{Molaverdikhani2020, Wardenier2025}, which is expected to result in emission spectra without strong line features.

The present study provides the first in-depth spectroscopic characterization of the UHJ TOI-2109b. We report detections of the spectral signatures of \ion{Fe}{i} and CO from the planet's dayside atmosphere, and the nondetection of line features in its nightside emission spectrum. In addition, we provide constraints on \mbox{TOI-2109b's} atmospheric parameters derived by a Bayesian retrieval framework. The planet was discovered via TESS photometry on a close orbit around a F-type star \citep{Wong2021} and confirmed through radial velocity (RV) measurements. With an extremely short orbital period of $P_\mathrm{orb}$\,$\sim$\,0.67\,d, an elevated dayside brightness temperature of $T_\mathrm{day}$\,$\sim$\,3630\,K, a particularly high planetary mass of $M_\mathrm{p}$\,$\sim$\,5.0\,$M_\mathrm{Jup}$ and density of $\rho_\mathrm{p}$\,$\sim$\,2.5\,g\,cm$^{-3}$, TOI-2109b represents a departure from the parameters typical of most UHJs. Beyond the discovery paper by \cite{Wong2021}, the literature on \mbox{TOI-2109b} remains limited to CHEOPS photometric observations by \cite{Harre2024}, which provide tentative evidence for orbital decay and suggest the possible presence of a nearby outer companion. TOI-2109b lies close to the lower boundary of an unexplored thermal regime in the range 3500\,K\,$<$\,$T_\mathrm{eq}$\,$<$\,4500\,K, which separates the hottest planet yet discovered -- KELT-9b \citep{Gaudi2017} -- from the currently known UHJs. While measurable abundances of a few molecular species have been found to persist in UHJ atmospheres with temperatures below this gap \citep[e.g.,][]{Yan2023, Cont2024, Pelletier2025}, no features of molecular lines have been detected at high spectral resolution in the atmosphere of KELT-9b as of March 2025\footnote{\url{https://research.iac.es/proyecto/exoatmospheres/view.php?name=KELT-9\%20b}}. This suggests that the temperature range between 3500\,K and 4500\,K could mark a transition in the thermochemical properties of UHJ atmospheres. As the first UHJ identified in this intermediate temperature range, \mbox{TOI-2109b} provides an important opportunity to investigate the connection between distinct thermochemical regimes. All parameters of the TOI-2109 system used in this work are summarized in the Table~\ref{table:planet-parameters}.

This paper is structured as follows. Sections~\ref{Observations} and \ref{Data reduction} describe the observations and data reduction procedures. The methodology for identifying the spectral emission lines from TOI-2109b's atmosphere, along with the resulting detections and nondetections, is presented in Sect.~\ref{Detection of the planetary emission lines}. Our retrieval framework and the derived atmospheric parameters are presented and discussed in Sect.~\ref{Retrieval of the atmospheric properties}. Finally, conclusions about our work are given in Sect.~\ref{Conclusions}.

\begin{table}
        \caption{Parameters of the \object{TOI-2109} (BD+16\,3058) system.}
        \label{table:planet-parameters}
        \centering 
        \renewcommand{\arraystretch}{1.4}
        \begin{threeparttable}
                \begin{tabular}{l@{\hspace{-6mm}}c}
                \hline\hline
                \noalign{\smallskip}
                Parameter & Value \\
                \noalign{\smallskip}
                \hline 
                \noalign{\smallskip}
                \textit{Planet} &  \\ 
                \noalign{\smallskip}
                Radius ($R_\mathrm{p}$) & $1.347 \pm 0.047$\,$R_\mathrm{Jup}$ \\
                Orbital period ($P_\mathrm{orb}$) & $0.67247414 \pm 0.00000028$\,d \\ 
                Transit epoch ($T_\mathrm{0}$) &  $2459378.459370 \pm 0.000059$\,BJD$_\mathrm{TDB}$ \\ 
                Orbital inclination ($i$) & $70.74 \pm 0.37$\,deg \\                         
                Surface gravity ($\log{g}$) & $3.836 \pm 0.071$\,cgs \\
                RV semi-amplitude ($K_\mathrm{p}$)      & $259.87 \pm 4.45$\,km\,s$^{-1}$ $^{(a)}$ \\                     
                Systemic velocity ($\varv_\mathrm{sys}$) & $-18.64_{-4.98}^{+5.86}$\,km\,s$^{-1}$ $^{(b)}$ \\
                \noalign{\smallskip} \hline \noalign{\smallskip}
                \textit{Star} &  \\  
                \noalign{\smallskip}
                Radius ($R_*$) & $1.698_{-0.057}^{+0.062}$\,$R_\mathrm{\sun}$ \\ 
                Mass ($M_*$) & $1.453 \pm 0.074$\,$M_\mathrm{\sun}$ \\ 
                Effective temperature ($T_\mathrm{eff}$) & $6540 \pm 160$\,K \\   
                
                \noalign{\smallskip}
                \hline                    
                \end{tabular}
                \vspace{0.3cm}
                \begin{tablenotes}
                \textbf{Notes.} $^{(a)}$ The RV semi-amplitude was calculated using the expression 
                \mbox{$K_\mathrm{p} = \left( 2 \pi G  M_* / P_\mathrm{orb} \right)^{1/3} \cdot \sin{i}$}, where $G$ is the gravitational constant, and all other parameters are listed in the table. $^{(b)}$ Value derived in this work, see Sect.~\ref{Dayside retrieval}; all other values from \cite{Wong2021}. 
                \end{tablenotes} 
        \end{threeparttable}
\end{table}

%

\section{Observations}
\label{Observations}

We observed the thermal emission signal of TOI-2109b over five visits at high spectral resolution using CARMENES \citep{Quirrenbach2018} and CRIRES$^+$ \citep{Dorn2023}. CARMENES is located at the 3.5-meter Calar Alto Telescope and consists of two fiber-fed high-resolution spectrograph channels that cover the wavelength ranges 520--960\,nm in the visible (VIS) and \mbox{960--1710\,nm} in the near-infrared (NIR) domain. The instrument operates at a spectral resolution of $R$\,$\sim$\,94\,600 in the VIS channel and $R$\,$\sim$\,80\,400 in the NIR channel, with 61 and 28 echelle orders, respectively. The CRIRES$^+$ instrument is a high-resolution slit spectrograph installed at the 8-meter Unit Telescope 3 of the Very Large Telescope. A total of 29 spectral settings provide full coverage of the wavelength interval \mbox{950--5380\,nm}. At the slit width of 0.2$^{\prime\prime}$ chosen for our observations, and under the assumption of homogeneous slit illumination, CRIRES$^+$ achieves a nominal resolving power of $R$\,$\sim$\,100\,000. For our observations, we selected the setting K2166, which provides wavelength coverage from 1921\,nm to 2472\,nm. This setting consists of seven echelle orders, each divided by two narrow gaps between the detectors, resulting in a total of 21 segments. Throughout this paper, for simplicity, we use the term ``wavelength segment'' to refer to both the spectral orders of CARMENES and the detector segments of CRIRES$^+$.

\begin{figure}
        \centering
        \includegraphics[width=\columnwidth]{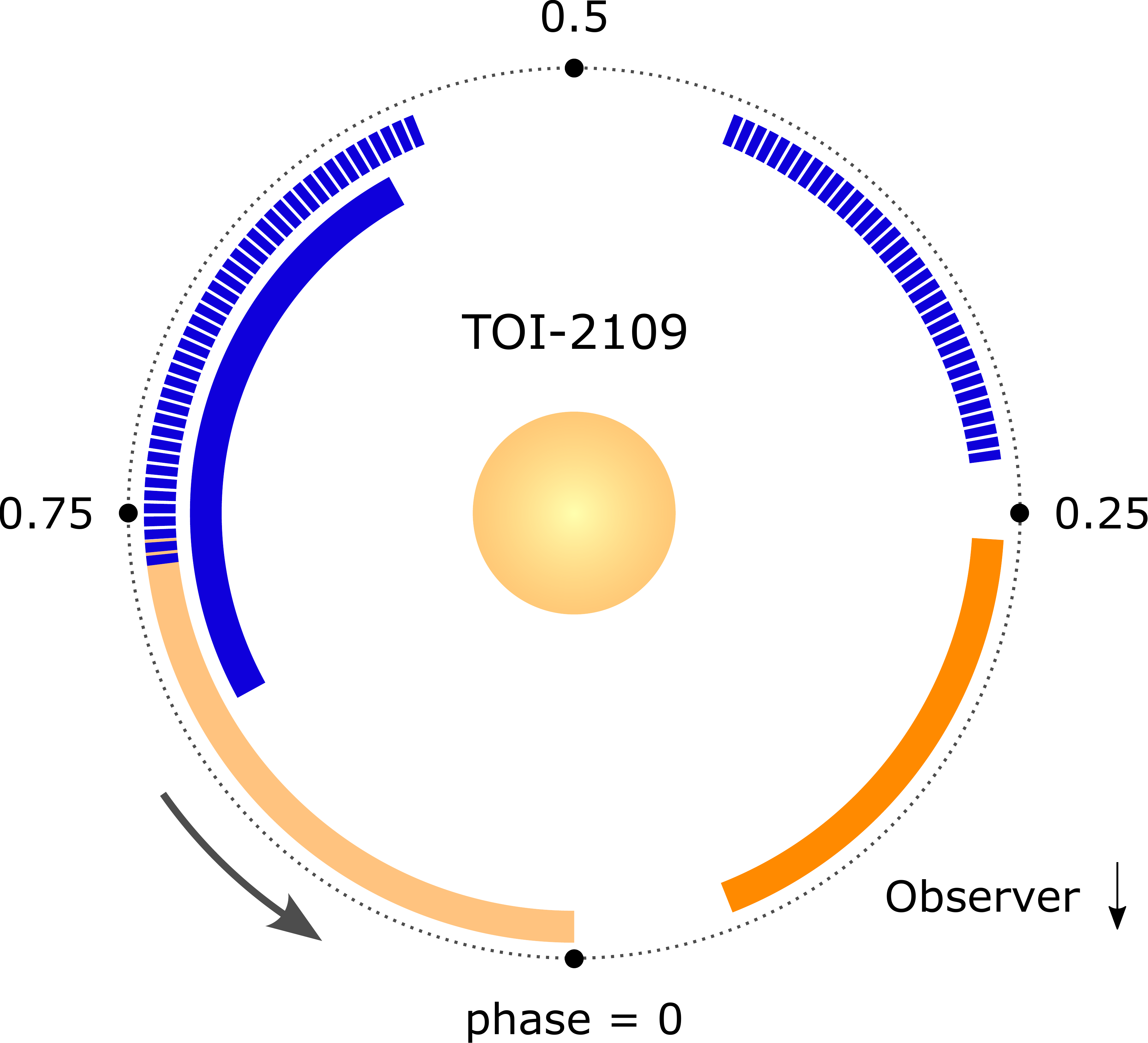}
        \caption{Orbital phase coverage of TOI-2109b observations. CARMENES observations are shown as hatched lines, while CRIRES$^+$ observations are shown as solid lines. Observations where the planet's day- and nightside are mostly aligned with the observer's line of sight are represented in blue and orange, respectively. The CRIRES$^+$ observation excluded due to poor data quality is shown in light orange.}
        \label{figure:phase-coverage}
\end{figure}

The acquired data cover the entire planetary orbit except for three small intervals during the primary and secondary eclipses and near the quadrature point at orbital phase 0.25. Therefore, our observations target the emission signals from both the day- and nightside hemispheres. Three spectral time series were taken at orbital phases when the dayside of the planet is mostly aligned with the observer's line of sight (CARMENES: pre- and post-eclipse; CRIRES$^+$: post-eclipse); another two CRIRES$^+$ observations were taken at orbital phases when the nightside of \mbox{TOI-2109b} is mostly facing the observer. The exposure time per spectral frame was 200\,s and 120\,s for CARMENES and CRIRES$^+$, respectively. An overview of the orbital phases covered is provided in Fig.~\ref{figure:phase-coverage} and further details of the observations are summarized in Table~\ref{table:obs-log}.

\begin{table*}
        \caption{Observation log.}
        \label{table:obs-log}
        \centering
        \renewcommand{\arraystretch}{1.4}
        \begin{threeparttable}
                \begin{tabular}{l c c c c c c c c }
                        \hline\hline
                        \noalign{\smallskip}
                        Instrument & Spectral resolution & Date & Phase coverage & Exposure time & Seeing & S/N & Airmass & $N_\mathrm{spectra}$  \\
                        \noalign{\smallskip}
                        \hline
                        \noalign{\smallskip}
                        CARMENES    & 94\,600 (VIS)     & 2022 May 22 & 0.56--0.77  & 200\,s & 1.5$^{\prime\prime}$ & 17 & 1.2--1.1 & 49\\ 
                                    & 80\,400 (NIR)     & 2022 May 22 & 0.56--0.77  & 206\,s & 1.5$^{\prime\prime}$ & 19 & 1.2--1.1 & 49\\
                                    & 94\,600 (VIS)     & 2022 June 7 & 0.27--0.44  & 200\,s & 1.4$^{\prime\prime}$ & 21 & 1.2--1.1 & 40\\ 
                                    & 80\,400 (NIR)     & 2022 June 7 & 0.27--0.44  & 206\,s & 1.4$^{\prime\prime}$ & 23 & 1.2--1.1 & 40\\ 
                        \noalign{\smallskip} \hline \noalign{\smallskip}
                        CRIRES$^+$  & 118\,900 $^{(a)}$     & 2024 April 8 & 0.58--0.83  & 120\,s & 0.7$^{\prime\prime}$ & 62 & 1.7--1.3--1.5 & 100\\ 
                                    & 143\,000 $^{(a)}$     & 2024 April 9 & 0.06--0.24  & 120\,s & 0.4$^{\prime\prime}$ & 73 & 1.7--1.3      & 76 \\ 
                                    & 100\,000 $^{(b)}$  & 2024 May 26 & 0.76--1.00  & 120\,s & 1.5$^{\prime\prime}$ & 38 & 1.7--1.3--1.4 & 102 \\              
                        \noalign{\smallskip}
                        \hline
                \end{tabular}
                \tablefoot{$^{(a)}$ The spectral resolution values for CRIRES$^+$ were derived from the width of the PSF; for the fiber-fed CARMENES spectrograph, the spectral resolution corresponds to the nominal instrument values. $^{(b)}$ The extended PSF homogeneously illuminated the slit, resulting in the instrument's nominal spectral resolution; the AO system failed for a total of 59 spectra, and due to poor data quality, the dataset was excluded from the analyses presented in this paper.}

        \end{threeparttable}      
\end{table*}

All observations were taken under good meteorological conditions, except for the CRIRES$^+$ night of 2024 May 26. The data from this night are affected by strong variability in the seeing conditions, which caused repeated failures of the adaptive optics (AO) system. We applied the data reduction and cross-correlation procedures described in Sects.~\ref{Data reduction} and \ref{Detection of the planetary emission lines} to this dataset. However, precise removal of the telluric contamination was prevented by strong temporal flux variations in the spectral time series caused by the repeated opening and closing of the AO loop. The residual telluric contamination resulted in the presence of artifacts in the cross-correlation analysis. We therefore concluded that the quality of this particular dataset was not suitable for drawing meaningful conclusions about the atmospheric conditions of TOI-2109b and excluded it from further analyses.

We used the reduction pipelines \texttt{caracal}\,v2.20 \citep{Zechmeister2014, Caballero2016} and \texttt{cr2res}\,1.4.4 to extract the order-by-order one-dimensional spectra and corresponding uncertainties from the raw frames\footnote{The extracted data, including the refined wavelength solution of CRIRES$^+$, are available at \url{https://zenodo.org/records/15037659} \citep{Lavail2025}.}. Both pipelines perform dark and flat field corrections, bad pixel removal, and provide a wavelength solution. The CRIRES$^+$ spectra were acquired using an ABBA nodding pattern, which consists of observing the target at two different slit positions (A and B), allowing the removal of sky background and detector artifacts during data extraction. We treated the A and B spectra as independent datasets throughout the data reduction steps in Sect.~\ref{Data reduction}, and combined their information in Sects.~\ref{Detection of the planetary emission lines} and \ref{Retrieval of the atmospheric properties}. 

For the CRIRES$^+$ observations, a precise wavelength solution is obtained only under the condition of a homogeneously illuminated spectrograph slit, which is not generally valid. Therefore, we performed a wavelength refinement step by applying \texttt{molecfit} \citep{Smette2015} to the time-averaged telluric absorption lines at the nodding positions A and B individually. A further correction of the wavelength solution for the individual exposures was not required, as time-dependent wavelength drifts determined by cross-correlation between the telluric lines were sufficiently small ($<0.15$\,pixels). In addition, for the CRIRES$^+$ data (except for the excluded night of 2024 May 26), the performance of the AO system was very good, resulting in a point spread function (PSF) smaller than the slit and thus a spectral resolution exceeding the nominal value of the spectrograph. We derived the specific spectral resolution of each observation from the width of the PSF. For a more detailed description of the wavelength correction procedures and the determination of the spectral resolution, we refer the reader to \cite{Cont2024}.

\section{Data reduction}
\label{Data reduction}

\subsection{Pre-processing the spectra}

Each observational dataset was pre-processed separately. For each wavelength segment, the extracted one-dimensional spectra were arranged chronologically in a two-dimensional array to form a spectral matrix. Variability in atmospheric conditions during the observations caused differences in the continuum level of the individual spectra. To correct for these differences, we normalized each individual spectrum to a common continuum using polynomial fitting (see \citealt{Cont2024}). We removed outliers in our data using Principal Component Analysis. With this approach, we modeled the spectral matrix (i.e., the two-dimensional array of flux as a function of wavelength and orbital phase), subtracted the model from the data, and identified pixels deviating by more than 5$\sigma$ in the resulting residuals. These pixels were then removed from the spectral matrix. Additionally, we masked wavelength regions with deep telluric lines, where the flux dropped below 50\% of the spectral continuum level.

\subsection{Removal of telluric and stellar lines}
\label{Removal of telluric and stellar lines}

Analyzing the planetary spectral signature requires the removal of telluric and stellar line contributions from our observations. To this end, we applied \texttt{SYSREM} to the pre-processed spectral matrix of each observation \citep{Tamuz2005}. Originally developed to identify and remove systematic effects in sets of stellar light curves, this algorithm has proven to be a powerful tool in the study of exoplanet atmospheres \citep[e.g.,][]{Birkby2013, Birkby2017}. When applied to high-resolution spectral time series, each wavelength bin of the spectral matrix is treated as an independent light curve. \texttt{SYSREM} iteratively derives a linear model in wavelength and time of the systematic contributions to the spectral matrix. These modeled systematics are subtracted from the data, yielding the so-called residual spectral matrix. We refer the reader to \cite{Czesla2024} for a detailed mathematical description of \texttt{SYSREM} in the context of exoplanet science.

The spectral matrix provided as input to the algorithm consists of the following main components: the stellar and telluric lines, the planetary spectral signature, and spectral noise. The positions of stellar and telluric features in wavelength space are stable over the duration of an observation and are therefore treated as systematic effects by \texttt{SYSREM}. Consequently, these components are efficiently modeled and removed from the spectral time series. In contrast, the spectral lines of the planet, which are buried in the noise, exhibit a continuously changing Doppler-shift caused by its orbital motion and are therefore largely unaffected by \texttt{SYSREM}.

\begin{figure}
        \centering
        \includegraphics[width=\columnwidth]{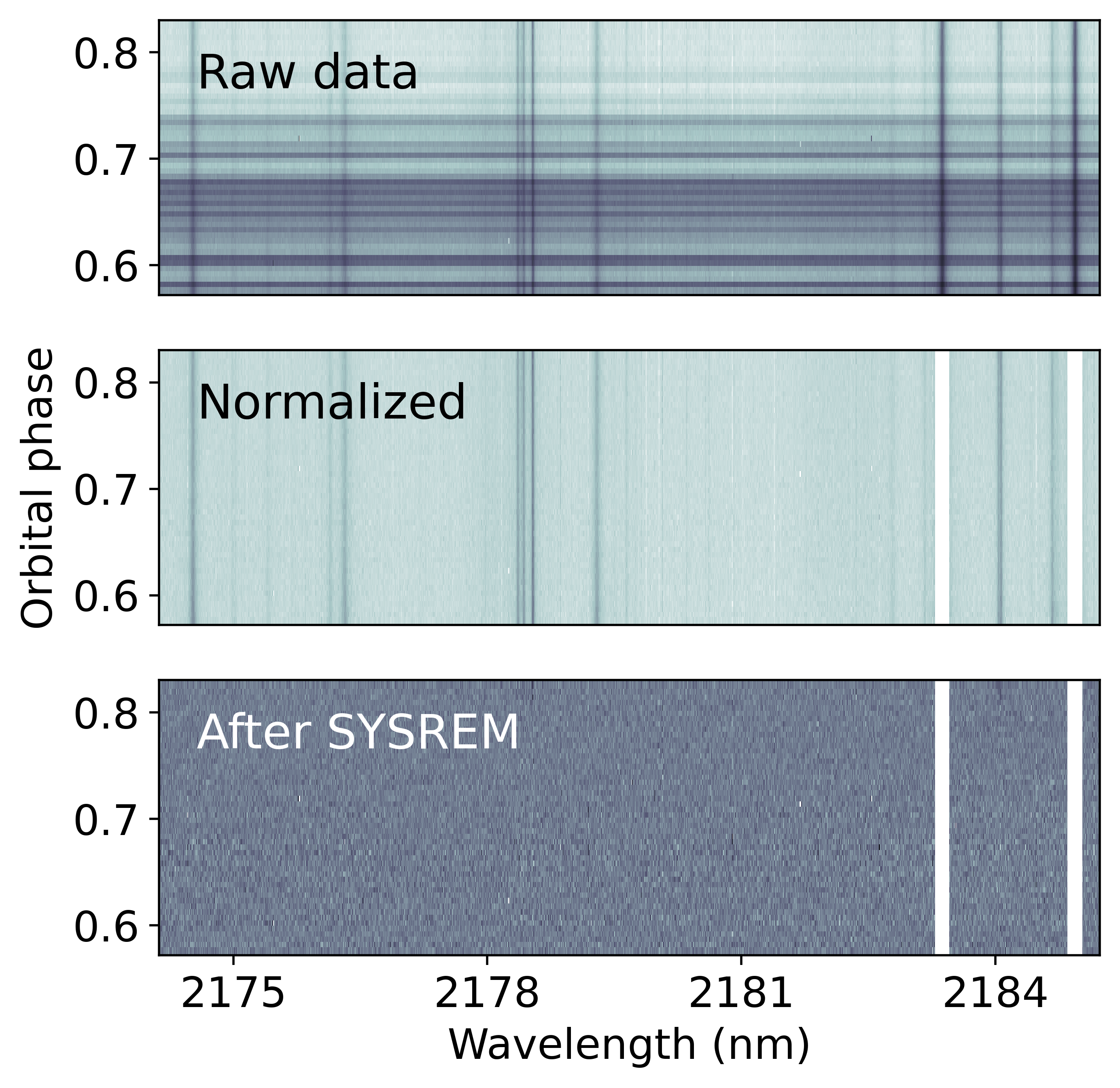}
        \caption{Example of data reduction steps for a selected CRIRES$^+$ wavelength range. \textit{Top panel}: Unprocessed spectra as extracted from the instrument pipeline. \textit{Middle panel}: Spectral matrix after continuum normalization and outlier correction; deep telluric lines are masked. \textit{Bottom panel}: Residual spectral matrix after telluric and stellar line removal with \texttt{SYSREM}.}
        \label{figure:data-reduction-steps}
\end{figure}

We followed the procedure described by \cite{Gibson2022}, which involves dividing the spectral matrix by the median spectrum before subtracting the systematics modeled by \texttt{SYSREM}. This approach preserves the relative strengths of the planetary spectral lines and allows us to account for distortions introduced by the algorithm when applying a Bayesian retrieval framework in Sect.~\ref{Retrieval of the atmospheric properties}. \texttt{SYSREM} was run for up to ten consecutive iterations, producing a residual spectral matrix for each iteration and wavelength segment. An overview of the data reduction process, including the use of \texttt{SYSREM}, is provided in Fig.\ref{figure:data-reduction-steps}.

%

\section{Detection of the planetary emission lines}
\label{Detection of the planetary emission lines}

We applied the cross-correlation method \citep[e.g.,][]{Snellen2010, Brogi2012, Birkby2013} to extract the faint spectral emission lines of TOI-2109b from the noise-dominated residual spectra. This method maps the information contained in the ensemble of planetary spectral lines onto a detectable signal peak by calculating the cross-correlation function (CCF) between an exoplanet model spectrum and the residual spectra. Typically, the spectral lines of the two chemical species \ion{Fe}{i} and CO dominate the emission signals from the dayside atmospheres of UHJs. The \ion{Fe}{i} lines are expected to be most prominent in the CARMENES wavelength range, while the CO lines are expected to be restricted to the CRIRES$^+$ wavelengths. Therefore, we search for \ion{Fe}{i} and CO in the CARMENES and CRIRES$^+$ dayside datasets, respectively (orbital phase coverage indicated in blue in Fig.~\ref{figure:phase-coverage}). Although the elevated dayside temperature of TOI-2109b suggests significant dissociation of H$_2$O, we also check for the corresponding emission lines, since this species, if present, can significantly affect the atmospheric metallicity and the carbon-to-oxygen (C/O) ratio in the retrieval described in Sect.~\ref{Retrieval of the atmospheric properties}. On the other hand, significant concentrations of CO and H$_2$O are expected in the cooler nightside atmospheres of UHJs. Consequently, we search for CO and H$_2$O in the nightside data of \mbox{TOI-2109b} (orbital phase coverage indicated in orange in Fig.~\ref{figure:phase-coverage}). For consistency with our dayside analysis, we also included \ion{Fe}{i} in our nightside investigations.

\subsection{Model spectra}
\label{Model spectra}

We modeled a planetary atmosphere consisting of 81 layers logarithmically spaced in pressure, ranging from $10^{-8}$\,bar to 1\,bar. The $T$-$p$ profile was parameterized with a low-pressure point ($T_1$, $p_1$) and a high-pressure point ($T_2$, $p_2$). An isothermal atmosphere was assumed at pressures below $p_1$ or higher than $p_2$. Between the two isothermal layers, a linear temperature pattern as a function of $\log{p}$ was adopted. The thermal conditions in TOI-2109b's atmosphere have not yet been studied in detail. Therefore, we used measurements of the atmospheric thermal conditions of the UHJs KELT-9b and \mbox{WASP-33b} to approximate the $T$-$p$ profile of TOI-2109b, which has an equilibrium temperature between those of the two planets. For our analysis of the planetary dayside, we employed a $T$-$p$ profile that lies between those determined by \cite{Kasper2021} and \cite{Cont2022b} for KELT-9b and WASP-33b, respectively. This $T$-$p$ profile corresponds to a thermal inversion and is defined by the pressure points $(T_1, p_1) = (5000\text{\,K}, 10^{-5}\text{\,bar})$ and $(T_2, p_2) = (3500\text{\,K}, 1\text{\,bar})$. To study the nightside atmosphere of TOI-2109b, we used a non-inverted $T$-$p$ profile similar to that retrieved by \cite{Mraz2024} for WASP-33b at orbital phases near the transit. The two-point parametrization of this profile is given by $(T_1, p_1) = (3000\text{\,K}, 10^{-2.5}\text{\,bar})$ and \mbox{$(T_2, p_2) = (4000\text{\,K}, 1\text{\,bar})$}. For both $T$-$p$ profiles used, the mean molecular weight and volume mixing ratios (VMRs) for the studied species were calculated with \texttt{FastChem} \citep{Stock2022}\footnote{\url{https://github.com/exoclime/FastChem}}, assuming equilibrium chemistry and solar elemental abundances. In addition, we computed a \mbox{$T$-$p$-wavelength} grid of the opacities for each chemical species to compute the corresponding planetary model spectrum. The opacities for \ion{Fe}{i} were obtained from the Kurucz line list \citep{Kurucz2018}, the CO line information was taken from \cite{Li2015}, and H$_2$O data were taken from the POKAZATEL line list \citep{Polyansky2018}. Eventually, the radiative transfer code \texttt{petitRADTRANS} \citep{Molliere2019} was used to generate the model spectra of the different chemical species.

\begin{figure*}
        \centering
        \includegraphics[width=\textwidth]{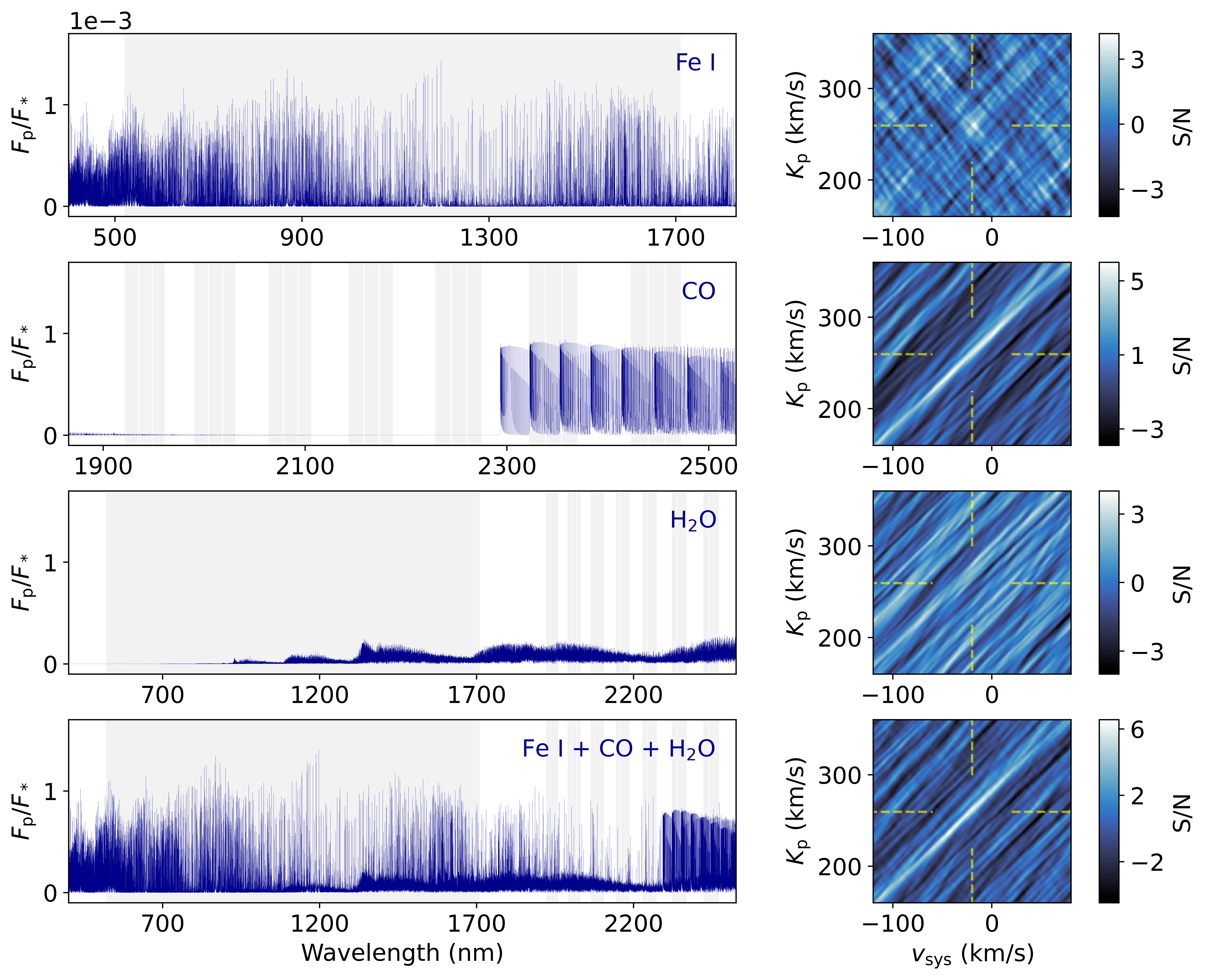}
        \caption{Spectral models (\textit{left panels}) and S/N maps (\textit{right panels}) of the chemical species investigated in the dayside atmosphere of TOI-2109b. The three \textit{top panels} show the information for the individual chemical species \ion{Fe}{i}, CO, and H$_2$O, the \textit{bottom panels} show the information for all species combined. We note that the species-combined signal is dominated by the CO detection with CRIRES$^+$. The \ion{Fe}{i} spectral signature mainly stems from CARMENES, the CO signal from the CRIRES$^+$ data; the investigated H$_2$O emission lines fall in the wavelength ranges of both instruments. The H$_2$O model spectrum shows particularly weak emission lines due to the predominant thermal dissociation of the species at the elevated temperatures of the TOI-2109b atmosphere. The wavelengths covered by CARMENES and the CRIRES$^+$ K2166 setting used in this work are indicated by the gray shaded area. The expected position of the planetary signal in the S/N maps is indicated by the yellow dashed lines. We show the S/N maps corresponding to the optimal \texttt{SYSREM} iteration number as determined in Sect.~\ref{Selection of SYSREM iterations}.}
        \label{figure:detections-dayside}
\end{figure*}

\begin{figure*}
        \centering
        \includegraphics[width=\textwidth]{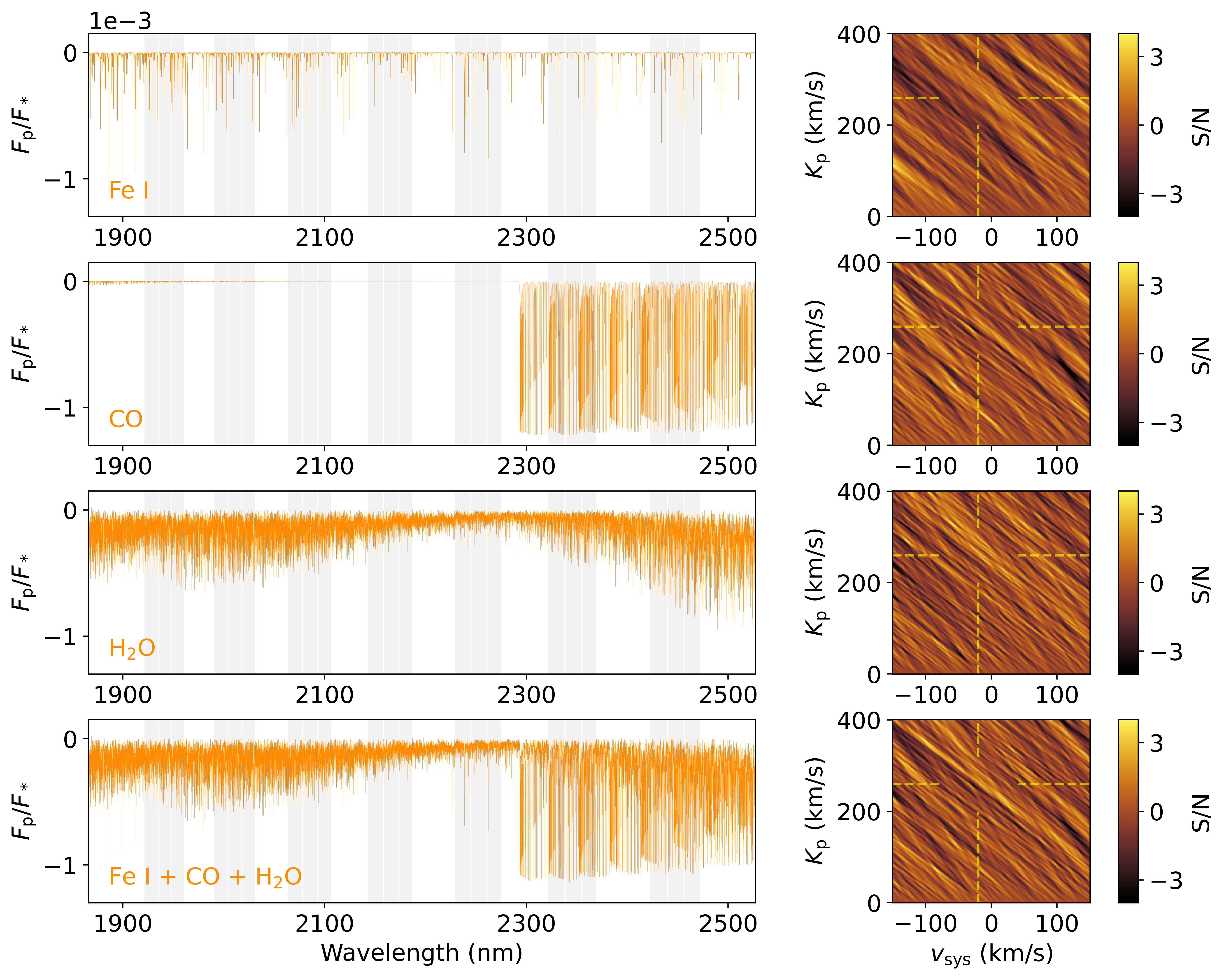}
        \caption{Same as Fig.~\ref{figure:detections-dayside} but for the nightside atmosphere of TOI-2109b. The investigated species result in nondetections. The nightside hemisphere was observed exclusively with CRIRES$^+$.}
        \label{figure:detections-nightside}
\end{figure*}

Since the pre-processed spectra are continuum normalized, normalization of the model spectra was also required. This step consisted of first calculating the planet-to-star flux ratio by dividing the model spectra by the blackbody spectrum of the host star, followed by normalization to the planetary continuum. As a last step, we convolved each model spectrum with the instrument profiles, resulting in the final spectra for cross-correlation. The normalized spectra for studying the day- and nightsides of TOI-2109b are shown in Figs.~\ref{figure:detections-dayside} and \ref{figure:detections-nightside}, respectively.

\subsection{Cross-correlation method}
\label{Cross-correlation method}

We applied the cross-correlation technique using the model spectra of the individual chemical species, as well as the species-combined model spectra. To this end, each model spectrum was shifted to a grid of Doppler-velocities between \mbox{--800\,km\,s$^{-1}$} and \mbox{+800\,km\,s$^{-1}$} with an increment of 1\,km\,s$^{-1}$ and interpolated to the wavelength solution of the data. For each velocity of the grid, the individual residual spectra were multiplied by the shifted model spectrum and weighted by the uncertainties, resulting in a so-called weighted CCF per exposure frame. For the residual spectrum number $i$ of the spectral time series, the CCF is thus defined as
\begin{equation}
    \mathrm{CCF}_i\left(\varv\right) = \sum_j \frac{ R_{ij} m_j\left(\varv\right) }{ {\sigma_{ij}}^2 },
\end{equation}
where $R_{ij}$ is the residual spectral matrix, $\sigma_{ij}$ are the uncertainties of the matrix, and $m_j$ is the spectral model Doppler-shifted by the velocity $\varv$. For each observational dataset and wavelength segment, the CCFs of the spectral time series were arranged in a two-dimensional array. The arrays of the different wavelength segments were then co-added, resulting in a CCF map for each dataset. Eventually, the CCF maps from different datasets were combined into a final CCF map. This was done by first merging the CCF maps of the individual datasets along the time axis and then sorting the rows of the resulting matrix according to the orbital phase. No weighting between different datasets was applied. This procedure allowed us to perform a combined analysis of the CARMENES datasets, as well as to merge the information from the A and B nodding positions of the CRIRES$^+$ data, which were treated independently throughout the data reduction procedures in Sect.~\ref{Data reduction}.

The CCF map was aligned to the rest frame of TOI-2109b over a range of orbital semi-amplitude velocity ($K_\mathrm{p}$) values. We assumed a circular orbit of the planet, with a RV described by
\begin{equation}
    \label{equation:planetary-rest-frame} 
    \varv_\mathrm{p} \left( t \right) = \varv_\mathrm{sys} + \varv_\mathrm{bary} \left( t \right) + K_\mathrm{p} \sin{2\pi\phi \left( t \right)},
\end{equation}
where $\varv_\mathrm{sys}$ is the velocity of the TOI-2109 system, $\varv_\mathrm{bary} \left( t \right)$ is the barycentric velocity correction, and $\phi \left( t \right)$ is the orbital phase. For each alignment, the CCF map was collapsed into a one-dimensional CCF by averaging over all orbital phases. The one-dimensional CCFs resulting from these alignments were further arranged into a two-dimensional array. Following \cite{Cont2024}, we then fit the distribution of all CCF values with a Gaussian function, divided the array by the resulting standard deviation, and obtained a signal-to-noise detection map (S/N map) as a function of the free parameters $K_\mathrm{p}$ and $\varv_\mathrm{sys}$. If the investigated spectral signature of the planet is present in the data, the S/N map will show a significant detection peak near the expected $K_\mathrm{p}$ and $\varv_\mathrm{sys}$ values.

\subsection{Number of \texttt{SYSREM} iterations}
\label{Selection of SYSREM iterations}

The number of \texttt{SYSREM} iterations applied to clean our data from telluric and stellar contamination can significantly affect the morphology of the S/N map, as well as the conclusions drawn by retrieval frameworks. Insufficient iterations result in incomplete correction of telluric and stellar residuals, biasing both the S/N map and the retrieved atmospheric parameters. On the other hand, using an excessive number of \texttt{SYSREM} iterations has the potential to not only remove telluric and stellar lines, but also to partially remove the planetary spectral signature. To objectively determine the optimal number of \texttt{SYSREM} iterations, we used the method proposed by \cite{Cheverall2023}, which is summarized as follows: first, a planetary model spectrum is Doppler-shifted using the expected $K_\mathrm{p}$ and $\varv_\mathrm{sys}$ values. The shifted spectral model is injected into the raw data, pre-processed, and cross-correlated with the unshifted planetary model spectrum. This procedure yields the injected cross-correlation function CCF$_\mathrm{inj}$. Subsequently, the differential cross-correlation function between the data with and without the injected spectral model \mbox{$\Delta \mathrm{CCF} = \mathrm{CCF}_\mathrm{inj} - \mathrm{CCF}$} is computed and converted into a S/N map for each iteration. The iteration corresponding to the highest significance in the S/N map is chosen as the optimal number of \texttt{SYSREM} iterations.

To derive the optimal \texttt{SYSREM} iteration number, we used the species-combined model spectra shown in the bottom left panels of Figs.~\ref{figure:detections-dayside} and \ref{figure:detections-nightside}. The recovered injected signal is maximized at iteration number six for CARMENES; three and seven consecutive iterations maximize the injected signal in the day- and nightside data of CRIRES$^+$, respectively. 

\begin{figure}
        \centering
        \includegraphics[width=0.5\textwidth]{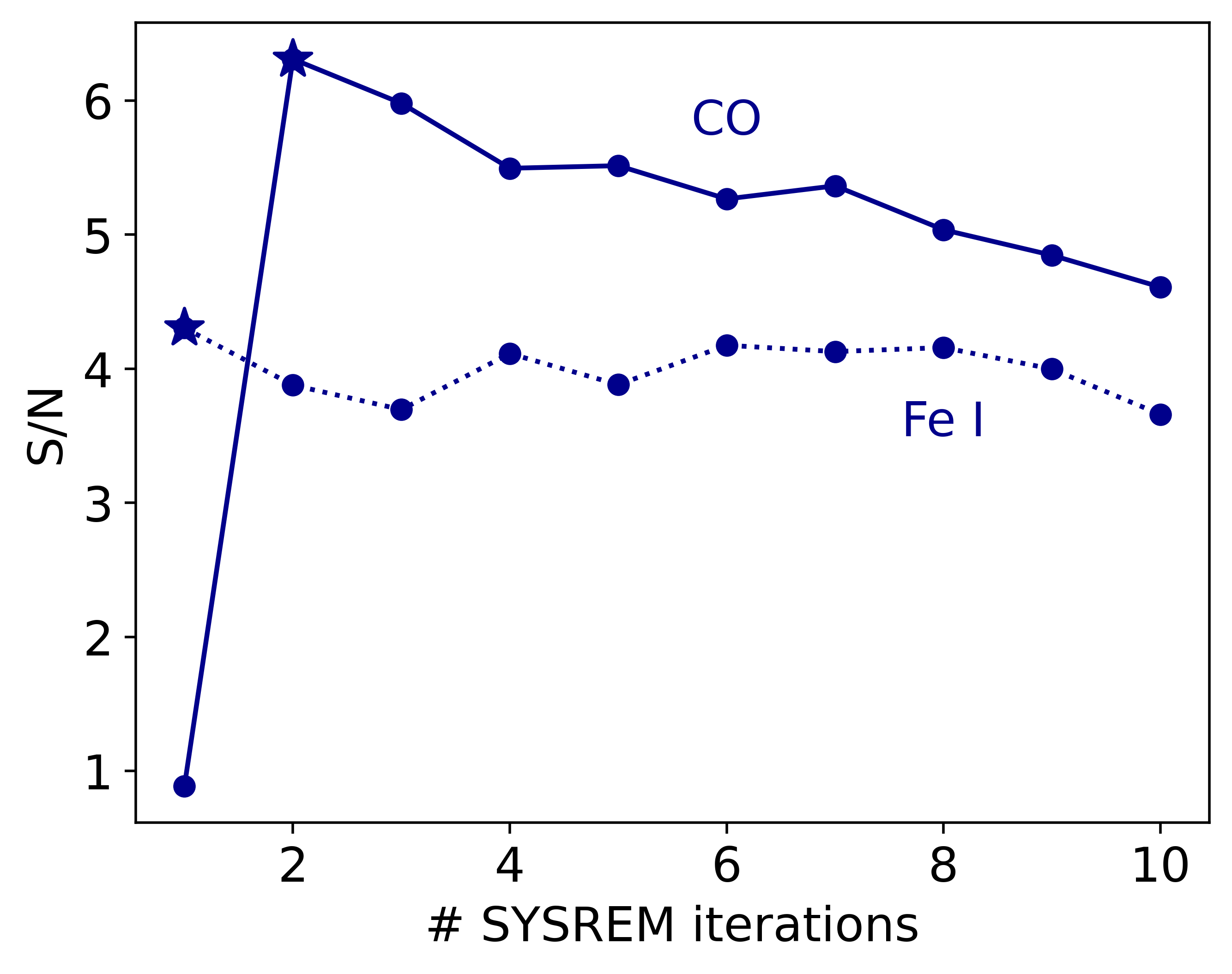}
        \caption{S/N values of \ion{Fe}{i} and CO as a function of \texttt{SYSREM} iterations. Iterations with the most significant S/N peaks are indicated by the star symbol.}
        \label{figure:sysrem-iterations}
\end{figure}

\subsection{Cross-correlation results}
\label{Cross-correlation results}

\subsubsection{Detection of the dayside signal}
\label{Detection of the dayside signal}

We detect spectral emission lines from the dayside atmosphere of TOI-2109b. In Fig.~\ref{figure:phase-coverage}, the orbital phase coverage of the dayside datasets is shown in blue color. Figure~\ref{figure:detections-dayside} presents the S/N maps resulting from the cross-correlation of our data with the \ion{Fe}{i}, CO, H$_2$O, and species-combined model spectra. The dashed lines indicate the expected location of the planetary signal in these maps. The expected $K_\mathrm{p}$ value is derived from the parameters in \cite{Wong2021}, while the value of $\varv_\mathrm{sys}$ is taken from the Gaia catalog \citep{Gaia2018, Gaia2023}.

We find tentative evidence for \ion{Fe}{i} emission lines consistent with the planetary rest frame in the CARMENES wavelength range. The signal is clearly visible at all \texttt{SYSREM} iterations, reaching its maximum strength at a S/N of 4.3 after one iteration. In addition, we identify a clear S/N detection peak of atmospheric CO close to the predicted values of $K_\mathrm{p}$ and $\varv_\mathrm{sys}$ in the CRIRES$^+$ data. The signal is visible at all {\tt SYSREM} iterations greater than one, with the strongest detection at S/N\,=\,6.3 after two consecutive iterations of the algorithm. No H$_2$O signal is detected at any \texttt{SYSREM} iteration. Cross-correlation with a model spectrum containing spectral lines from all three chemical species and combining data from both instruments gives a maximum detection at S/N\,=\,6.6 after two iterations of \texttt{SYSREM}. Comparison of this detection strength with that of the individual species shows that the spectral signature of CO, whose lines are restricted to the CRIRES$^+$ wavelength range, clearly dominates the species-combined signal. This is a reasonable result considering the significantly higher photon collecting power of the \mbox{8-meter} mirror at the Very Large Telescope available to CRIRES$^+$ compared to the 3.5-meter mirror at Calar Alto Observatory used by CARMENES. The evolution of the signal strengths as a function of {\tt SYSREM} iterations is shown in Fig.~\ref{figure:sysrem-iterations}. The detection of spectral lines in emission unambiguously proves the presence of an atmospheric temperature inversion on the dayside of TOI-2109b, which is in agreement with theoretical work on the temperature structure of UHJ atmospheres \citep[e.g.,][]{Hubeny2003, Fortney2008}. In addition to the CARMENES data, we cross-correlated our \ion{Fe}{i} model spectrum with the CRIRES$^+$ dayside dataset. However, this resulted in a nondetection, likely due to the limited number and low intensity of \ion{Fe}{i} spectral emission lines in the wavelength range of the selected CRIRES$^+$ setting. 

While most high-resolution studies have shown a strong increase in detection significance between the first run of \texttt{SYSREM} and subsequent iterations, the \ion{Fe}{i} signal shows a stable pattern that includes the first iteration. This behavior can be attributed to two factors. First, the planetary \ion{Fe}{i} lines are predominantly located at the blue end of the CARMENES wavelength range, where the telluric contamination is minimal. Second, the normalization of the spectral time series with the median spectrum prior to the \texttt{SYSREM} application in Sect.~\ref{Removal of telluric and stellar lines} already provides a preliminary removal of telluric and stellar lines. As a result, telluric and stellar contamination, and consequently \texttt{SYSREM}, have a limited effect on the \ion{Fe}{i} signal.

In addition to \ion{Fe}{i}, previous studies with CARMENES have detected several other chemical species in the atmospheres of UHJs, including \ion{Ti}{i}, \ion{V}{i}, \ion{Si}{i}, OH, and TiO \citep[e.g.,][]{Landman2021, Cont2021, Cont2022b, Cont2022a, Ridden-Harper2023, Guo2024}. Therefore, we also checked for these species in the CARMENES data of TOI-2109b, but found no significant detections. The nondetections are not surprising, since these species typically produce weaker signals than \ion{Fe}{i}, whose S/N is relatively low in our analysis. Thus, even if these species are present in TOI-2109b's atmosphere, their signals are unlikely to overcome the noise in our data. For Ti- and V-bearing species, depletion mechanisms such as gravitational settling or cold trapping on the nightside may provide an additional explanation for their nondetection \citep[e.g.,][]{Spiegel2009, Parmentier2013}.

\subsubsection{Nondetection of the nightside signal}
\label{Nondetection of the nightside signal}

No significant emission signal is detected in the CRIRES$^+$ data taken when the nightside atmosphere of TOI-2109b is facing the observer. The nightside observations correspond to the orbital phases indicated in orange color in Fig.~\ref{figure:phase-coverage}. The S/N maps of the nondetections obtained from cross-correlation with the \ion{Fe}{i}, CO, H$_2$O, and the species-combined model spectra are reported in Fig.~\ref{figure:detections-nightside}. Our nondetections are most likely due to an approximately isothermal nightside $T$-$p$ profile in the pressure range probed by high-resolution spectroscopy, rather than the absence of the investigated species. In general, an isothermal structure results in a featureless blackbody spectrum of the planetary atmosphere, which prevents the detection of a cross-correlation signal. In a broader context, our nondetections are not surprising, given that only a small number of tentative nightside signals have been found for other gas giant exoplanets at high spectral resolution \citep[i.e., HD~179949b, WASP-33b;][]{Matthews2024, Yang2024, Mraz2024}.

Previous work by \cite{Wong2021} suggests that, depending on the thermal conditions, the nightside hemisphere of \mbox{TOI-2109b} may exhibit detectable emission features of CO and H$_2$O. Their study focused on spectral signals accessible with low-resolution space-based spectroscopy, which is primarily sensitive to pressure levels near the planetary photosphere. In addition to the atmospheric layers at lower pressures, high-resolution ground-based spectroscopy can also probe the pressure range near the photosphere and is therefore potentially sensitive to these spectral features. Our nondetection of molecular signals suggests that TOI-2109b has nightside thermal conditions resulting in an emission signal without strong spectral features. According to Fig.~16 of \cite{Wong2021}, this may be the case for nightside temperatures below 2500\,K, which is consistent with the temperature constraints obtained from our nightside retrieval in Sect.~\ref{Nightside retrieval}.

%

\section{Retrieval of the atmospheric properties}
\label{Retrieval of the atmospheric properties}

\subsection{Retrieval framework}
\label{Retrieval framework}

The retrieval of TOI-2109b's atmospheric properties employs a framework closely resembling that used by \cite{Cont2024} for characterizing the atmosphere of WASP-178b. Our implementation relies on the {\tt petitRADTRANS} radiative transfer code \citep{Molliere2019} to forward model the planetary emission spectrum. The atmosphere was modeled with 25 layers, uniformly distributed on a logarithmic pressure scale from $10^{-8}$\,bar to 1\,bar. We included the opacities of the chemical species studied in Sect.~\ref{Cross-correlation results}, i.e., \ion{Fe}{i}, CO, and H$_2$O. In addition, the presence of H$^-$ was considered for both hemispheres, as this species is a critical continuum opacity source that can mute emission features in the spectra of UHJs \citep[e.g.,][]{Arcangeli2018, Lothringer2018}. The thermal and chemical structure of the planetary atmosphere was parametrized by a two-point \mbox{$T$-$p$} profile with the high-pressure point $p_2$ defined as \mbox{$\log{p_2} = \log{p_1} + \mathrm{d}p$} and $\mathrm{d}p$ restricted to positive values, and by the atmospheric metallicity ([M/H]) and C/O ratio. Following \cite{Lesjak2025}, we assumed that most atmospheric metals vary with the overall metallicity and hence \mbox{[M/H]\,=\,[Fe/H]\,=\,[O/H]}, while the carbon abundance is determined by the C/O ratio and oxygen abundance. Using these parameters, the VMRs of the different chemical species were calculated with the \texttt{FastChem} code \citep{Stock2022}. We incorporated the effect of rotational broadening into the calculation of the spectral model, which is parameterized by the planet's equatorial rotation velocity $\varv_\mathrm{eq}$ \citep{Diaz2011}. The resulting model spectrum was converted to the planet-to-star flux ratio and convolved with the instrumental profile as described in Sect.~\ref{Model spectra}.

For each individual spectrum of our high-resolution time series, we used $K_\mathrm{p}$ and $\varv_\mathrm{sys}$ to Doppler-shift the spectral model to the planetary rest frame and interpolated to the wavelength solution of the observational data. The short orbital period of TOI-2109b causes its RV to change significantly during a single spectral exposure, smearing the observed spectral signal over multiple detector pixels. We accounted for this effect by convolving each Doppler-shifted model spectrum in velocity space with a box-function kernel corresponding to the RV change during each individual exposure. These procedures resulted in a two-dimensional representation of the model spectrum with the same shape as the residual spectral matrix. We adopted the filtering method described by \cite{Gibson2022} to account for the presence of potential distortion effects introduced into our data by \texttt{SYSREM}.

High-resolution spectroscopy lacks the information from the spectral continuum level that is lost during the telluric and stellar line correction with \texttt{SYSREM}. Therefore, only the relative strengths of the spectral lines with respect to the local continuum are measured. On the other hand, photometric and low-resolution spectroscopy observations can measure absolute flux levels that encode the spectral continuum information. In addition to our high-resolution spectroscopy datasets, photometric measurements of TOI-2109b are available in the literature. \cite{Wong2021} measured the secondary eclipse depth of \mbox{TOI-2109b} using space- and ground-based photometry with TESS and the Palomar/WIRC instrument, obtaining eclipse depth values of 731$\pm$46\,ppm and 2012$\pm$80\,ppm, respectively. From the nightside, they obtained a photometric flux of 9$\pm$43\,ppm in the TESS passband. To incorporate this information into our retrieval framework, we modeled the photometric data points with \texttt{petitRADTRANS}. As the TESS bandpass covers a substantial fraction of the visible wavelength range, which can encompass emission features of a variety of atomic metals, their oxides, and hydrides, we included the opacities of Fe, Ti, Na, K, Ca, Mg, TiO, VO, FeH, CrH, and CaH into our photometric modeling. In addition, we included the CO and H$_2$O opacities, which are expected to have the most dominant effect on the WIRC \mbox{K band} photometry. The contribution from the continuum opacity source H$^-$ was also taken into account. We applied the same thermal and chemical parametrization as for the high-resolution spectroscopy data. The resulting emission flux was integrated over the bandpass of each instrument. Since the integration of the model spectrum occurs over relatively wide bandpasses, the {\tt petitRADTRANS} low-resolution mode was used to speed up the radiative transfer calculations. The opacities used in the low-resolution mode are available in the \texttt{petitRADTRANS} opacity database \footnote{\url{https://petitradtrans.readthedocs.io/en/latest/content/available_opacities.html}}.

Our approach follows that of \cite{Yan2022a}, assuming that the dayside flux observed in photometry results solely from thermal emission, with no contribution from reflected stellar light. This approximation is motivated by the extremely high temperatures of TOI-2109b's dayside atmosphere, which prevent the formation of reflective clouds or hazes \citep[e.g.,][]{Lothringer2018, Kitzmann2018}. The assumption is also supported by measurements of the reflection properties of other close-in gas giant exoplanets, which indicate very low albedo values \citep[e.g.,][]{Bell2017, Shporer2019}.

Following the pioneering works of \cite{Brogi-Line2019} and \cite{Gibson2020}, a standard Gaussian log likelihood function was used to compare the model spectra to the observations. For the high-resolution spectroscopy data this function is defined as
\begin{equation}
    \ln{\mathcal{L}} = -\frac{1}{2}\sum_{i,j} \left[ \frac{\left(R_{ij} - M_{ij}\right)^2}{\left(\beta \sigma_{ij}\right)^2} + \ln{2 \pi \left(\beta \sigma_{ij}\right)^2} \right],
\end{equation}
where $R_{ij}$ and $M_{ij}$ are the individual elements of the residual spectral matrix and the filtered spectral forward model, respectively. The uncertainties of the residual spectra are denoted by $\sigma_{ij}$; $\beta$ acts as a scaling factor to correct for possible over- or underestimation of the uncertainties. For each high-resolution spectroscopy dataset, the log likelihood functions resulting from the individual wavelength segments were coadded. This yielded a total of five log likelihood functions for the analysis of the planetary dayside (two CARMENES VIS, two CARMENES NIR, and one CRIRES$^+$ dataset) and one function for the analysis of the planetary nightside (one CRIRES$^+$ dataset). The residual spectra used were those corresponding to the optimal number of \texttt{SYSREM} iterations computed in Sect.~\ref{Selection of SYSREM iterations}.

The Gaussian log likelihood function for the photometric data points was defined as
\begin{equation}
    \ln{\mathcal{L}_\mathrm{ph}} = -\frac{1}{2} \left[ \frac{(R_\mathrm{ph} - \delta_\mathrm{ph})^2}{\sigma_\mathrm{ph}^2} + \ln{2 \pi \sigma_\mathrm{ph}^2} \right].
\end{equation}
In this expression, $R_\mathrm{ph}$ is the photometric flux, $\delta_\mathrm{ph}$ is the modeled flux, and $\sigma_\mathrm{ph}$ is the measurement uncertainty. We computed the log likelihood functions for both the TESS and Palomar/WIRC data points.

Eventually, the log likelihood functions were combined into a final log likelihood function for the planetary dayside and a separate final log likelihood function for the planetary nightside. This combination was achieved by coadding the log likelihood functions of the high-resolution spectroscopy and photometric measurements separately for the day- and nightside. To obtain the posterior parameter estimates, we explored the resulting posterior probability distribution, which is proportional to the likelihood for uniform priors, by Markov Chain Monte Carlo (MCMC) sampling using the \texttt{emcee} software package \citep{Foreman-Mackey2013}. In summary, our high-resolution retrieval framework includes the following parameters: the temperature profile parameters $T_1$, $T_2$, $p_1$, d$p$; the chemical properties represented by [M/H] and the C/O ratio; the equatorial rotation velocity $\varv_\mathrm{eq}$; the velocity parameters $K_\mathrm{p}$ and $\varv_\mathrm{sys}$; the noise scaling parameters of the high-resolution spectrographs $\beta_\mathrm{CV}$, $\beta_\mathrm{CN}$, $\beta_\mathrm{C+}$ (the instruments have independent noise scaling factors, where CV, CN, and C+ denote CARMENES VIS, CARMENES NIR, and CRIRES$^+$, respectively). For each free parameter, 32 walkers with 30\,000 steps each were used in the MCMC sampling process.

\subsection{Retrieval results and discussion}

\subsubsection{Dayside retrieval}
\label{Dayside retrieval}

Our retrieval is able to constrain physical and chemical properties in the dayside atmosphere of TOI-2109b. The corner plot in Fig.~\ref{figure:corner-plot-dayside} shows the posterior distributions along with the correlations of the atmospheric parameters. A summary of the best-fit retrieval parameters is provided in Table~\ref{table:retrieval-results}. 

\begin{table*}
        \caption{Results of atmospheric retrievals on TOI-2109b.
        }
        \label{table:retrieval-results}
        \centering
        \renewcommand{\arraystretch}{1.5}
        \begin{threeparttable}
                \begin{tabular}{l c c c c c}
                        \hline\hline
                        \noalign{\smallskip}
                        Parameter & Dayside retrieval & Nightside retrieval  & Prior dayside & Prior nightside & Unit  \\
                        \noalign{\smallskip}
                        \hline
                        \noalign{\smallskip}
                        $T_1$                  & $4601_{-609}^{+1557}$         & $<8159$                   & (2000, 9000)  & (100, 9000)     &    K \\ 
                        $T_2$                  & $3173_{-727}^{+300}$          & $<2364$                   & (1000, 7000)  & (100, 7000)     &    K \\ 
                        $\log{p_1}$            & $-4.92_{-1.89}^{+1.99}$       & $< -3.04$                 & (-8, 0)       & (-8, 0)         &    $\log{\mathrm{bar}}$\\         
                        $\mathrm{d}p$          & $>2.37$                       & $< 5.86$                  & (0, 8)        & (0, 8)          &    $\log{\mathrm{bar}}$\\    
                        $\log{p_2}$ $^{(a)}$   & $0.35_{-1.48}^{+2.44}$        & $-3.51_{-1.86}^{+2.17}$   & \ldots        & \ldots          &    $\log{\mathrm{bar}}$\\    
                        $\mathrm{[M/H]}$       & $0.36_{-1.17}^{+1.42}$        & \ldots                    & (-3, 3)       & 0               &    dex \\ 
                        $\mathrm{C/O}$         & $>0.15$                       & \ldots                    & (0, 1.5)      & 0.55            &    \ldots\\ 
                        $\varv_\mathrm{eq}$    & $12.07_{-5.78}^{+5.75}$       & \ldots                    & (0, 30)       & 10              &    km\,s$^{-1}$ \\ 
                        $K_\mathrm{p}$         & $262.95_{-5.72}^{+6.65}$      & \ldots                    & (190, 330)    & 260             &    km\,s$^{-1}$\\ 
                        $\varv_\mathrm{sys}$   & $-18.64_{-4.98}^{+5.86}$      & \ldots                    & (-90, 30)     & $-19$           &    km\,s$^{-1}$ \\ 
                        \noalign{\smallskip}
                        \hline
                        \noalign{\smallskip}                    
                        $\beta$ CARMENES VIS   & $0.9082 \pm 0.0002$           & \ldots                    & (0, 3)        & \ldots          &    \ldots\\  
                        $\beta$ CARMENES NIR   & $0.7403 \pm 0.0003$           & \ldots                    & (0, 3)        & \ldots          &    \ldots\\     
                        $\beta$ CRIRES$^+$     & $1.7016 \pm 0.0007$           & $1.4049 \pm 0.0007$       & (0, 3)        & (0, 3)          &    \ldots\\     
                        
                        \noalign{\smallskip}
                        \hline
                \end{tabular}
                \tablefoot{$^{(a)}$ This table also includes $\log{p_2}$, which is not treated as a free parameter in the retrieval; $\log{p_2}$ is calculated as the sum of the free parameters $\log{p_1}$ and $\mathrm{d}p$.}
        \end{threeparttable}   
\end{table*}

\begin{figure*}
        \centering
        \includegraphics[width=\textwidth]{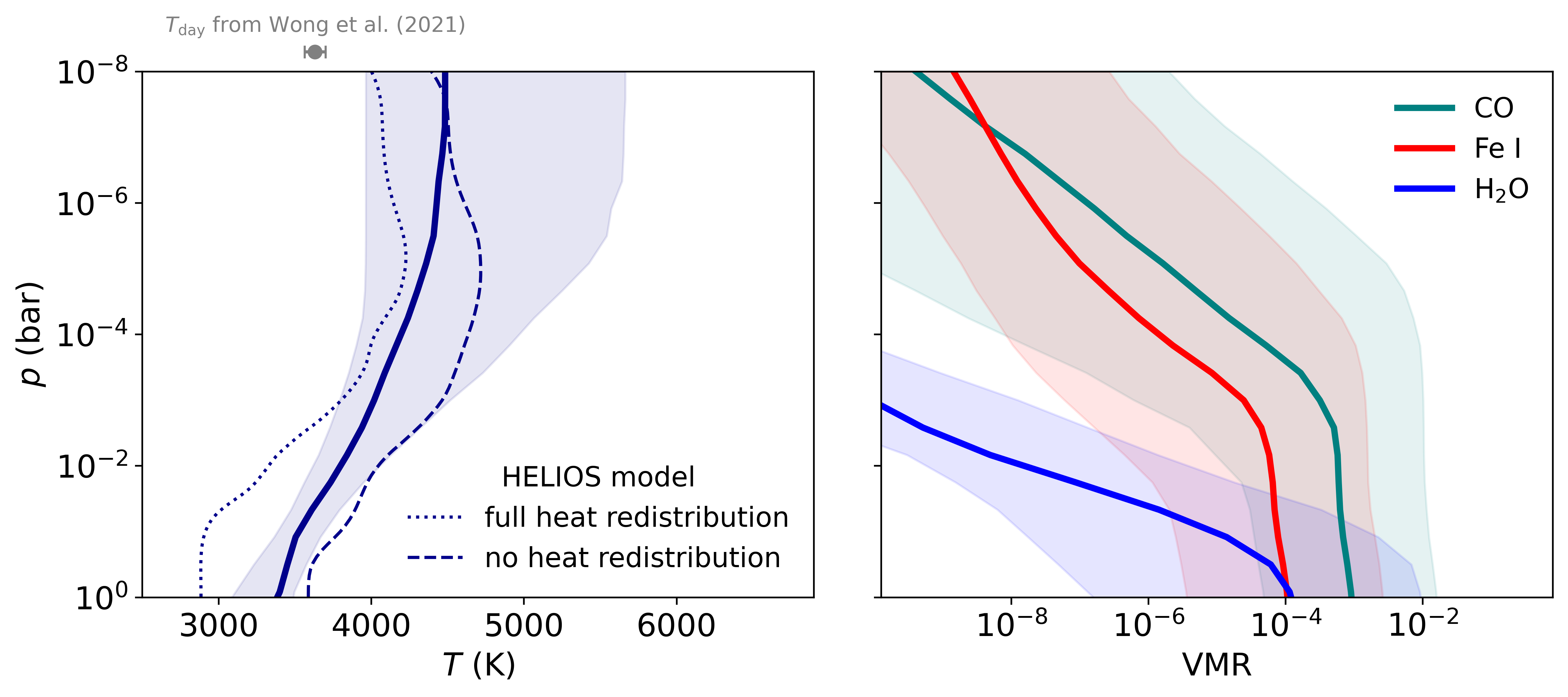}
        \caption{Atmospheric temperature and abundance profiles. The \textit{left panel} shows a comparison between the median $T$-$p$ profile from the retrieval (solid line) and the $T$-$p$ profiles from the self-consistent \texttt{HELIOS} model. The \texttt{HELIOS} temperature curves represent the two limiting cases of no and full day- to nightside heat redistribution. The uncertainty interval of the retrieved $T$-$p$ curve is represented by the blue shaded area. Our $T$-$p$ measurement is slightly closer to the no heat redistribution scenario than to the full heat redistribution scenario.         
        For comparison, we also include the dayside temperature measurement from \cite{Wong2021}, shown as the gray data point. This measurement is expected to reflect the thermal conditions at low atmospheric altitudes, and is therefore in line with the retrieved $T$-$p$ profile. Since \cite{Wong2021} do not provide a specific pressure value, we show this data point outside the $T$-$p$ space. 
        The \textit{right panel} shows the abundances (solid lines) of the chemical species studied by cross-correlation and the corresponding uncertainties (shaded areas). Both \ion{Fe}{i} and CO abundances drop significantly at pressures below $\sim$\,$10^{-3}$\,bar, indicating that our data are most sensitive to atmospheric layers at or above this pressure level.}
        \label{figure:dayside-Tp-VMRs}
\end{figure*}

The retrieved $T$-$p$ profile is shown in Fig.~\ref{figure:dayside-Tp-VMRs}, confirming the presence of the thermal inversion layer detected in Sect.~\ref{Cross-correlation method}. This finding is consistent with theoretical predictions \citep[e.g.,][]{Hubeny2003, Fortney2008, Arcangeli2018} and is in line with previous observational studies of UHJs \citep[e.g.,][]{Nugroho2017, Cont2022a, Yan2022b, Lesjak2025}. The thermal inversion spans a temperature range between approximately 3200\,K and 4600\,K in the lower and upper planetary atmosphere, respectively. Our retrieval provides narrower constraints on the thermal conditions at high pressure levels compared to low pressure levels. The smaller uncertainties at higher pressures result from the inclusion of photometric data points, which predominantly probe the layers at low atmospheric altitudes. Therefore, the measured temperature of these atmospheric layers is expected to be close to the value of $3631\pm69$\,K determined by \cite{Wong2021} exclusively from TESS and Palomar/WIRC photometric measurements. We note that the temperature derived by our retrieval is slightly lower, with a discrepancy of approximately 100\,K between the confidence intervals of their measurement and that of our work. We attribute this offset to differences in the modeling approaches used: while \cite{Wong2021} assume a blackbody emission spectrum to determine TOI-2109b's dayside temperature, our analysis includes the additional contribution of line emission, which requires a lower temperature to obtain the same band-integrated flux. On the other hand, the larger uncertainties of the $T$-$p$ point describing the high-altitude atmospheric layers are caused by the predominant ionization and dissociation of the studied chemical species in the low-pressure regime.

To compare with the $T$-$p$ profile obtained from our retrieval, we calculated the thermal structure of TOI-2109b's dayside atmosphere using the \texttt{HELIOS}\footnote{\url{https://github.com/exoclime/HELIOS}} radiative-convective equilibrium code \citep{Malik2017, Malik2019}. We followed the procedures described by \cite{Fossati2021} with the only difference that we include cross sections of all molecules available in the DACE\footnote{\url{https://dace.unige.ch/dashboard}} database. The atomic line opacities also included \mbox{\ion{C}{i-ii}}, \ion{O}{i-ii}, \ion{Na}{i-ii}, \ion{Mg}{i-ii}, \ion{Si}{i-ii}, \ion{K}{i-ii}, \ion{Ca}{i-ii}, \ion{Ti}{i-ii}, \ion{Cr}{i-ii}, and \ion{Fe}{i-ii}, which we pre-tabulated using the \texttt{HELIOS-K}\footnote{\url{https://github.com/exoclime/HELIOS-K}} package \citep{GrimmHeng2015} and the original Kurucz\footnote{\url{http://kurucz.harvard.edu/linelists/gfall/}} line list \citep{Kurucz2018}. Since \texttt{HELIOS} does not include photochemistry, we assumed equilibrium abundances which we obtained with the \texttt{FastChem} code \citep{Stock2022}.

Our \texttt{HELIOS} calculations considered two extreme scenarios: no heat redistribution and full heat redistribution from the dayside to the nightside atmosphere of the planet. The $T$-$p$ curve from our retrieval lies between the two \texttt{HELIOS} models, which is a reasonable result since the expected heat redistribution should fall between the two extremes. We also note that the gradient of the retrieved $T$-$p$ curve and that from the theoretical models are in agreement over the full pressure range of the thermal inversion layer. Figure~\ref{figure:dayside-Tp-VMRs} shows that the \texttt{HELIOS} model without heat redistribution is slightly closer to our $T$-$p$ measurement than the full redistribution scenario, suggesting that TOI-2109b's atmosphere may have a limited heat redistribution efficiency. However, further observations to refine the constraints on the dayside temperature profile are needed to allow a conclusive characterization of the planetary heat redistribution properties.

We derived an atmospheric metallicity of [M/H]\,=\,$0.36_{-1.17}^{+1.42}$\,dex, which is consistent with the host star value of $-0.07_{-0.06}^{+0.07}$\,dex \citep{Wong2021}. The atmospheric C/O ratio is only weakly constrained by our retrieval, excluding values below approximately 0.15. However, between the lower limit of 0.15 and a value of about 0.9, the posterior density estimate shows a maximum consistent with solar values. We attribute the relatively broad [M/H] interval and the inability of our retrieval to robustly constrain the C/O ratio to a degeneracy between the two parameters, as well as to degeneracies with the $T$-$p$ profile. The degeneracies can be identified as diagonal patterns in the correlation plots shown in Fig.~\ref{figure:corner-plot-dayside}. In principle, stronger detections of \ion{Fe}{i} and CO would be able to mitigate these degeneracies. However, our signals of these species appear to be insufficient for this purpose. This is in line with previous studies that were able to constrain [M/H] and the C/O ratio with significantly stronger detections than those achieved in our work \citep[e.g.,][]{Line2021}.

The VMRs of \ion{Fe}{i}, CO, and H$_2$O derived from the $T$-$p$ and chemical retrieval parameters are shown in Fig.~\ref{figure:dayside-Tp-VMRs}. Despite the relatively weak constraints on [M/H] and the C/O ratio, our retrieval yields VMRs in the lower atmosphere of approximately $10^{-4}$ for \ion{Fe}{i} and $10^{-3}$ for CO, with uncertainties of $\pm$\,1.5\,dex. Towards higher atmospheric altitudes the VMRs decrease significantly and the uncertainties increase. This effect can be attributed to increasing ionization and dissociation of \ion{Fe}{i} and CO, and suggests that our observations are mainly sensitive to the atmospheric layers at higher pressures. Significant VMRs of H$_2$O are obtained only at the highest pressures studied, which are typically probed exclusively by photometric measurements. The rapid decrease in H$_2$O abundances toward lower pressures is consistent with our nondetection of H$_2$O with high-resolution spectroscopy in Sect.~\ref{Detection of the planetary emission lines}.

Our retrieval yields relatively tight confidence intervals for the planetary velocity parameters encoded in the Doppler-shift of the spectral lines. The orbital and systemic velocity parameters are derived as $K_\mathrm{p}$\,=\,$262.95_{-5.72}^{+6.65}$\,km\,s$^{-1}$ and $\varv_\mathrm{sys}$\,=\,$-18.64_{-4.98}^{+5.86}$\,km\,s$^{-1}$, respectively. The retrieved $K_\mathrm{p}$ is consistent with the value predicted from the orbital period and adopted stellar mass. However, direct comparison of our $\varv_\mathrm{sys}$ measurement with literature values is difficult due to inconsistencies between previous systemic velocity measurements. For instance, \cite{Wong2021} report \mbox{$-25.64 \pm 0.11$\,km\,s$^{-1}$} and \mbox{$-25.61 \pm 0.21$\,km\,s$^{-1}$} with the TRES and FIES spectrographs, respectively, which differ from Gaia measurements of \mbox{$-19. 72 \pm 3.46$\,km\,s$^{-1}$} and \mbox{$-22.28 \pm 1.23$}\,km\,s$^{-1}$ \citep{Gaia2018, Gaia2023}. Consequently, we treat the $\varv_\mathrm{sys}$ value derived in our work as an independent measurement of the \mbox{TOI-2109} systemic velocity.

We observe significant spectral line broadening, which, assuming it is caused by planetary rotation, allows us to derive an effective rotational velocity of $\varv_\mathrm{eq}$\,=\,$12.07_{-5.78}^{+5.75}$\,km\,s$^{-1}$. This value is consistent with the velocity of $\sim$\,10\,km\,s$^{-1}$ that is derived under the assumption that the rotation of TOI-2109b is tidally locked. The absence of a significant velocity excess in the retrieved $\varv_\mathrm{eq}$ value suggests the lack of strong atmospheric dynamics, as phenomena such as super-rotating winds or turbulent flow would produce additional line broadening beyond what is expected from planetary rotation. This finding is in line with the picture of inefficient atmospheric heat redistribution suggested by the retrieved $T$-$p$ profile. We caution, however, that drawing firm conclusions about the impact of atmospheric dynamics on TOI-2109b's heat redistribution efficiency is difficult given the relatively large uncertainties in $\varv_\mathrm{eq}$, and that further interpretation would require additional observations.

The noise scaling terms $\beta_\mathrm{CV}$ and $\beta_\mathrm{CN}$ are close to one, indicating a proper estimate of the CARMENES uncertainties. The CRIRES$^+$ data show a slight underestimation of the uncertainties, resulting in a noise scaling term $\beta_\mathrm{C+}$ greater than one. This discrepancy, which was not observed in previous studies with the CRIRES$^+$ instrument \citep[e.g.,][]{Cont2024, Lesjak2025}, can be attributed to changes in the noise estimation routine between versions of the \texttt{cr2res} extraction pipeline\footnote{see Sect.~6.1.3 of the \texttt{cr2res} user manual at \url{https://ftp.eso.org/pub/dfs/pipelines/instruments/cr2res/cr2re-pipeline-manual-1.4.4.pdf}}. Pipeline versions prior to \texttt{cr2res}\,1.4.0 used a more conservative approach based on residual noise, while later versions used Poisson noise, resulting in a more optimistic estimation of the uncertainties.

In the context of the broader UHJ population, our results suggest that the atmospheric thermochemical regime of \mbox{TOI-2109b} represents a transition between the properties of UHJs below the thermal gap and those of the extreme object KELT-9b. The presence of significant CO abundances contrasts with the chemical properties of \mbox{KELT-9b}, for which no CO molecular features have been detected in the planetary dayside. On the other hand, the chemistry of TOI-2109b's dayside atmosphere also differs from that of the UHJs with temperatures cooler than 3500\,K. This is because H$_2$O, which has been detected in the emission spectra of several UHJs -- including WASP-33b ($T_\mathrm{eq}$\,$\sim$\,3250\,K; \citealt{Nugroho2021, Finnerty2023}), the third hottest known UHJ -- is largely absent from the dayside of TOI-2109b. Finally, the evidence for inefficient day-night heat transport suggests that \mbox{TOI-2109b} has heat redistribution properties more similar to those of UHJs with temperatures below the thermal gap than to those of KELT-9b.

\subsubsection{Nightside retrieval}
\label{Nightside retrieval}

In the retrieval analysis targeting the nightside atmosphere of TOI-2109b, we set the chemical, velocity, and broadening parameters to fixed values due to their poor constraints when left as free parameters. The chemical parameters [M/H] and C/O ratio were set to solar values, consistent with the results obtained for the planetary dayside. The $K_\mathrm{p}$ and $\varv_\mathrm{sys}$ values were fixed to those listed in Table~\ref{table:planet-parameters}, and the spectral lines were broadened according to a tidally locked planetary rotation with \mbox{$\varv_\mathrm{eq}$\,=\,10\,km\,s$^{-1}$}. This setup allowed us to focus on investigating the $T$-$p$ profile of the planetary nightside atmosphere.

\begin{figure}
        \centering
        \includegraphics[width=\columnwidth]{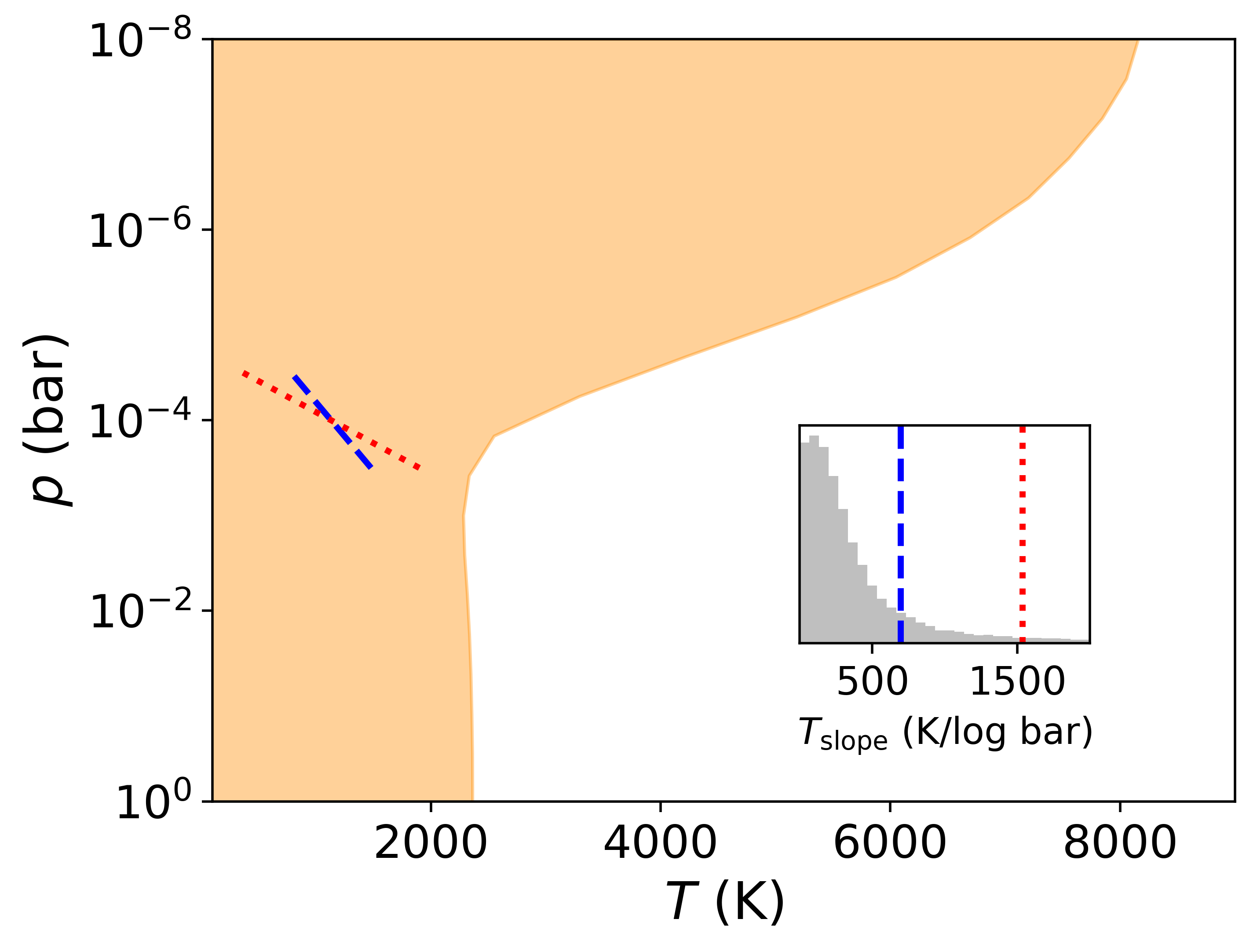}
        \caption{Upper temperature limits in the nightside hemisphere of TOI-2109b and posterior distribution of the atmospheric temperature gradient. The orange shaded area represents the 95th percentile confidence interval of the temperatures as a function of atmospheric pressure. The \textit{inset panel} shows the posterior distribution of the $T$-$p$ gradient for thermal profiles with decreasing temperatures towards lower pressures. The 84th and 95th percentile confidence intervals at $695$\,K/$\log{\mathrm{bar}}$ and $1535$\,K/$\log{\mathrm{bar}}$ are indicated by the blue dashed and red dotted lines, respectively. These gradients represent upper limits below which the atmospheric signal cannot be retrieved in the present data. The gradients are shown as blue dashed and red dotted lines in $T$-$p$ space.}
        \label{figure:nightside-Tp-VMRs}
\end{figure}

The posterior distributions and correlations between the atmospheric parameters are illustrated in the corner plot in Fig.~\ref{figure:corner-plot-nightside}, showing that only upper limits on the nightside temperatures can be retrieved. The upper limits of the nightside \mbox{$T$-$p$} profile over the entire pressure range studied are shown in Fig.~\ref{figure:nightside-Tp-VMRs}. At pressures greater than $10^{-4}$\,bar, the atmospheric temperature is restricted to values below approximately 2400\,K. This result is consistent with the maximum nightside temperature of $2500$\,K previously derived by \cite{Wong2021} from TESS photometric measurements. On the other hand, the atmospheric temperature at pressures below $10^{-4}$\,bar in TOI-2109b's nightside remains largely unconstrained. This suggests a limited ability of our retrieval to distinguish between different $T$-$p$ profile scenarios in this pressure range, presumably due to the absence of a sufficiently strong spectral signal. In fact, both an isothermal \mbox{$T$-$p$} profile and a \mbox{$T$-$p$} profile exhibiting thermal inversion, where temperatures at low pressures reach values high enough to cause ionization and dissociation of the species studied, produce signals with no spectral lines, which would be consistent with our nondetection. In principle, our retrieval is expected to be sensitive only to $T$-$p$ profiles with decreasing temperatures towards the upper layers of the atmosphere. This is because the limited impact of ionization and dissociation on the studied chemical species allows for the formation of significant spectral lines in this scenario. However, we are unable to constrain such a $T$-$p$ profile, and therefore rule out the existence of a strong thermal gradient towards lower temperatures in the upper atmosphere of TOI-2109b.

To place upper limits on the thermal gradient below which the atmospheric signal cannot be retrieved in the present data, we analyzed the posterior distribution of temperature gradients, defined as
\begin{equation}
T_\mathrm{slope} = \frac{T_1 - T_2}{\log{p_1} - \log{p_2}},
\end{equation}
using only those MCMC samples corresponding to a \mbox{$T$-$p$} profile decreasing with altitude. 
The probability distribution in Fig.~\ref{figure:nightside-Tp-VMRs} peaks at low $T_\mathrm{slope}$ values, with an extended tail towards higher values. This tail corresponds to thermal profiles with rapid temperature decreases over a narrow pressure range, resulting in isothermal profiles over almost the entire pressure range. The tail is a consequence of our two-point $T$-$p$ parametrization, which also allows for sharp temperature changes, which are likely unphysical. As a result, the 84th percentile confidence interval at $695$\,K/$\log{\mathrm{bar}}$ provides a reliable upper limit of the $T$-$p$ gradient in TOI-2109b's atmosphere, while we consider the 95th percentile interval commonly used for this purpose to be biased by the tail of the distribution.

We note that an alternative explanation for the weak constraints on the nightside thermal structure of TOI-2109b is that the $T$-$p$ profile may be more complex than a purely inverted, non-inverted, or isothermal pattern. Such a profile could produce spectral features that combine both emission and absorption components (for an example, see Fig.~2 of \citealt{Schwarz2015}). In this scenario, our two-point $T$-$p$ parameterization, which only allows for a single temperature gradient, would be insufficient to accurately describe the thermal structure, thereby limiting the ability of the retrieval to constrain the atmospheric conditions. The use of more flexible $T$-$p$ parameterizations in future studies may allow a more effective characterization of exoplanet nightside conditions \citep[e.g.][]{Pelletier2021, Gandhi2023}.

%

\section{Conclusions}
\label{Conclusions}

We have characterized the atmospheric properties of the UHJ TOI-2109b using high-resolution spectroscopy with CARMENES and CRIRES$^+$ at visible and near-infrared wavelengths. Applying the cross-correlation technique and a Bayesian retrieval framework incorporating broadband photometric data, we were able to derive the following key results:

\begin{enumerate}

    \item We identified emission lines of \ion{Fe}{i} and CO in the dayside spectrum of TOI-2109b. The detected Doppler-shifts of the spectral lines are consistent with the planet's orbital motion. Emission lines of H$_2$O remain undetected, most likely due to thermal dissociation of this species at the extreme temperatures of the planetary dayside (Sect.~\ref{Detection of the dayside signal});

    \item Our finding of \ion{Fe}{i} and CO spectral lines in emission proves the presence of a thermal inversion layer in the dayside atmosphere, which is consistent with the expectations from theoretical work on UHJs (Sect.~\ref{Detection of the dayside signal}). A joint retrieval with secondary eclipse depth measurements from TESS and Palomar/WIRC allowed us to obtain tight constraints on the $T$-$p$ profile. Weaker constraints on the metallicity and C/O ratio are found due to a degeneracy between the two parameters; the retrieved values are consistent with solar abundances (Sect.~\ref{Dayside retrieval});

    \item Comparison of the retrieved $T$-$p$ profile with self-consistent models suggests an inefficient day- to nightside heat transport in TOI-2109b's atmosphere. This finding is supported by a broadening of the spectral emission lines consistent with a tidally locked rotation, indicating the absence of strong day- to nightside wind flows (Sect.~\ref{Dayside retrieval});

    \item Finally, no significant emission signal was detected in the nightside observations of TOI-2109b (Sect.~\ref{Nondetection of the nightside signal}). A retrieval combining CRIRES$^+$ high-resolution spectroscopy and TESS photometry provides only upper limits on the gradient of the $T$-$p$ profile and temperature in the nightside atmosphere (Sect.~\ref{Nightside retrieval}). 
    
\end{enumerate}

The present study provides the first in-depth atmospheric characterization of TOI-2109b, a UHJ located at the lower end of the unexplored thermal regime between 3500\,K and 4500\,K. This gap separates the well studied sample of UHJs with lower temperatures from the extreme case of KELT-9b, the hottest exoplanet orbiting a main sequence star known to date. Overall, our findings of prominent CO spectral emission, absence of H$_2$O features, and evidence for low day-night heat transport suggest that the chemical and thermal properties of TOI-2109b correspond to a transition between the properties of the UHJs below the thermal gap and KELT-9b. Future observations of TOI-2109b will help to further connect the unique physical and chemical properties of the extreme object KELT-9b to the population of gas giants at lower temperatures.

%

\begin{acknowledgements}
    CRIRES$^+$ is an ESO upgrade project carried out by Th\"{u}ringer Landessternwarte Tautenburg, Georg-August Universit\"{a}t G\"{o}ttingen, and Uppsala University. The project is funded by the Federal Ministry of Education and Research (Germany) through Grants 05A11MG3, 05A14MG4, 05A17MG2 and the Knut and Alice Wallenberg Foundation. This project is based on observations collected at the European Organisation for Astronomical Research in the Southern Hemisphere under the ESO programmes 113.26GE.001 and 113.26GE.002. 
    CARMENES is an instrument at the Centro Astron\'omico Hispano en Andaluc{\'i}a (CAHA) at Calar Alto (Almer\'{\i}a, Spain), operated jointly by the Junta de Andaluc\'ia and the Instituto de Astrof\'isica de Andaluc\'ia (CSIC). CARMENES was funded by the Max-Planck-Gesellschaft (MPG), the Consejo Superior de Investigaciones Cient\'{\i}ficas (CSIC), the Ministerio de Econom\'ia y Competitividad (MINECO) and the European Regional Development Fund (ERDF) through projects FICTS-2011-02, ICTS-2017-07-CAHA-4, and CAHA16-CE-3978, and the members of the CARMENES Consortium (Max-Planck-Institut f\"ur Astronomie, Instituto de Astrof\'{\i}sica de Andaluc\'{\i}a, Landessternwarte K\"onigstuhl, Institut de Ci\`encies de l'Espai, Institut f\"ur Astrophysik G\"ottingen, Universidad Complutense de Madrid, Th\"uringer Landessternwarte Tautenburg, Instituto de Astrof\'{\i}sica de Canarias, Hamburger Sternwarte, Centro de Astrobiolog\'{\i}a and Centro Astron\'omico Hispano-Alem\'an), with additional contributions by the MINECO, the Deutsche Forschungsgemeinschaft (DFG) through the Major Research Instrumentation Programme and Research Unit FOR2544 ``Blue Planets around Red Stars'', the Klaus Tschira Stiftung, the states of Baden-W\"urttemberg and Niedersachsen, and by the Junta de Andaluc\'{\i}a. 
    We acknowledge support from DFG through Germany's Excellence Strategy EXC 2094 -- 390783311, the priority program SPP 1992 ``Exploring the Diversity of Extrasolar Planets'' DFG~PR~36~24602/41, and project 31466515; 
    the LMU-Munich Fraunhofer-Schwarzschild Fellowship; 
    the Knut and Alice Wallenberg Foundation through grant 2018.0192; 
    the National Natural Science Foundation of China through project 42375118; 
    the Agencia Estatal de Investigaci\'on (AEI/10.13039/501100011033) of the Ministerio de Ciencia e Innovaci\'on and the ERDF ``A way of making Europe'' through projects 
    PID2022-137241NB-C4[1:4], 
    PID2021-125627OB-C31, 
    PID2021-126365NB-C21, 
    RYC2022-037854-I, 
    and the Centre of Excellence ``Severo Ochoa'' and ``Mar\'ia de Maeztu'' 
    awards to the Instituto de Astrof\'isica de Canarias (CEX2019-000920-S), Instituto de Astrof\'isica de Andaluc\'ia (CEX2021-001131-S) and Institut de Ci\`encies de l'Espai (CEX2020-001058-M). 
    This work has made use of the following Python packages: {\tt Astropy} \citep{AstropyCollaboration2013}, {\tt CMasher} \citep{vanderVelden2020}, {\tt corner} \citep{ForemanMackey2016}, {\tt GNU Parallel} \citep{Tange2021}, {\tt Matplotlib} \citep{Hunter2007}, {\tt NumPy} \citep{Harris2020}, {\tt PyAstronomy} \citep{Czesla2019}, and {\tt SciPy} \citep{Virtanen2020}.
\end{acknowledgements}


\bibliographystyle{aa} 
\bibliography{references}

\begin{thebibliography}{104}
\expandafter\ifx\csname natexlab\endcsname\relax\def\natexlab#1{#1}\fi

\bibitem[{{Arcangeli} {et~al.}(2018){Arcangeli}, {D{\'e}sert}, {Line}, {Bean}, {Parmentier}, {Stevenson}, {Kreidberg}, {Fortney}, {Mansfield}, \& {Showman}}]{Arcangeli2018}
{Arcangeli}, J., {D{\'e}sert}, J.-M., {Line}, M.~R., {et~al.} 2018, \apjl, 855, L30

\bibitem[{{Astropy Collaboration} {et~al.}(2013){Astropy Collaboration}, {Robitaille}, {Tollerud}, {Greenfield}, {Droettboom}, {Bray}, {Aldcroft}, {Davis}, {Ginsburg}, {Price-Whelan}, {Kerzendorf}, {Conley}, {Crighton}, {Barbary}, {Muna}, {Ferguson}, {Grollier}, {Parikh}, {Nair}, {Unther}, {Deil}, {Woillez}, {Conseil}, {Kramer}, {Turner}, {Singer}, {Fox}, {Weaver}, {Zabalza}, {Edwards}, {Azalee Bostroem}, {Burke}, {Casey}, {Crawford}, {Dencheva}, {Ely}, {Jenness}, {Labrie}, {Lim}, {Pierfederici}, {Pontzen}, {Ptak}, {Refsdal}, {Servillat}, \& {Streicher}}]{AstropyCollaboration2013}
{Astropy Collaboration}, {Robitaille}, T.~P., {Tollerud}, E.~J., {et~al.} 2013, \aap, 558, A33

\bibitem[{{Bell} \& {Cowan}(2018)}]{BellCowan2018}
{Bell}, T.~J. \& {Cowan}, N.~B. 2018, \apjl, 857, L20

\bibitem[{{Bell} {et~al.}(2017){Bell}, {Nikolov}, {Cowan}, {Barstow}, {Barman}, {Crossfield}, {Gibson}, {Evans}, {Sing}, {Knutson}, {Kataria}, {Lothringer}, {Benneke}, \& {Schwartz}}]{Bell2017}
{Bell}, T.~J., {Nikolov}, N., {Cowan}, N.~B., {et~al.} 2017, \apjl, 847, L2

\bibitem[{{Bello-Arufe} {et~al.}(2022){Bello-Arufe}, {Cabot}, {Mendon{\c{c}}a}, {Buchhave}, \& {Rathcke}}]{Bello-Arufe2022a}
{Bello-Arufe}, A., {Cabot}, S. H.~C., {Mendon{\c{c}}a}, J.~M., {Buchhave}, L.~A., \& {Rathcke}, A.~D. 2022, \aj, 163, 96

\bibitem[{{Birkby} {et~al.}(2013){Birkby}, {de Kok}, {Brogi}, {de Mooij}, {Schwarz}, {Albrecht}, \& {Snellen}}]{Birkby2013}
{Birkby}, J.~L., {de Kok}, R.~J., {Brogi}, M., {et~al.} 2013, \mnras, 436, L35

\bibitem[{{Birkby} {et~al.}(2017){Birkby}, {de Kok}, {Brogi}, {Schwarz}, \& {Snellen}}]{Birkby2017}
{Birkby}, J.~L., {de Kok}, R.~J., {Brogi}, M., {Schwarz}, H., \& {Snellen}, I.~A.~G. 2017, \aj, 153, 138

\bibitem[{{Borsato} {et~al.}(2023){Borsato}, {Hoeijmakers}, {Prinoth}, {Thorsbro}, {Forsberg}, {Kitzmann}, {Jones}, \& {Heng}}]{Borsato2023a}
{Borsato}, N.~W., {Hoeijmakers}, H.~J., {Prinoth}, B., {et~al.} 2023, \aap, 673, A158

\bibitem[{{Brogi} {et~al.}(2023){Brogi}, {Emeka-Okafor}, {Line}, {Gandhi}, {Pino}, {Kempton}, {Rauscher}, {Parmentier}, {Bean}, {Mace}, {Cowan}, {Shkolnik}, {Wardenier}, {Mansfield}, {Welbanks}, {Smith}, {Fortney}, {Birkby}, {Zalesky}, {Dang}, {Patience}, \& {D{\'e}sert}}]{Brogi2023}
{Brogi}, M., {Emeka-Okafor}, V., {Line}, M.~R., {et~al.} 2023, \aj, 165, 91

\bibitem[{{Brogi} \& {Line}(2019)}]{Brogi-Line2019}
{Brogi}, M. \& {Line}, M.~R. 2019, \aj, 157, 114

\bibitem[{{Brogi} {et~al.}(2012){Brogi}, {Snellen}, {de Kok}, {Albrecht}, {Birkby}, \& {de Mooij}}]{Brogi2012}
{Brogi}, M., {Snellen}, I. A.~G., {de Kok}, R.~J., {et~al.} 2012, \nat, 486, 502

\bibitem[{{Caballero} {et~al.}(2016){Caballero}, {Gu{\`a}rdia}, {L{\'o}pez del Fresno}, {Zechmeister}, {de Juan}, {Alonso-Floriano}, {Amado}, {Colom{\'e}}, {Cort{\'e}s-Contreras}, {Garc{\'{\i}}a-Piquer}, {Gesa}, {de Guindos}, {Hagen}, {Helmling}, {Hern{\'a}ndez Casta{\~n}o}, {K{\"u}rster}, {L{\'o}pez-Santiago}, {Montes}, {Morales Mu{\~n}oz}, {Pavlov}, {Quirrenbach}, {Reiners}, {Ribas}, {Seifert}, \& {Solano}}]{Caballero2016}
{Caballero}, J.~A., {Gu{\`a}rdia}, J., {L{\'o}pez del Fresno}, M., {et~al.} 2016, in \procspie, Vol. 9910, Observatory Operations: Strategies, Processes, and Systems VI, 99100E

\bibitem[{{Cheverall} {et~al.}(2023){Cheverall}, {Madhusudhan}, \& {Holmberg}}]{Cheverall2023}
{Cheverall}, C.~J., {Madhusudhan}, N., \& {Holmberg}, M. 2023, \mnras, 522, 661

\bibitem[{{Cont} {et~al.}(2024){Cont}, {Nortmann}, {Yan}, {Lesjak}, {Czesla}, {Lavail}, {Reiners}, {Piskunov}, {Hatzes}, {Boldt-Christmas}, {Kochukhov}, {Marquart}, {Nagel}, {Rains}, {Rengel}, {Seemann}, \& {Shulyak}}]{Cont2024}
{Cont}, D., {Nortmann}, L., {Yan}, F., {et~al.} 2024, \aap, 688, A206

\bibitem[{{Cont} {et~al.}(2021){Cont}, {Yan}, {Reiners}, {Casasayas-Barris}, {Molli{\`e}re}, {Pall{\'e}}, {Henning}, {Nortmann}, {Stangret}, {Czesla}, {L{\'o}pez-Puertas}, {S{\'a}nchez-L{\'o}pez}, {Rodler}, {Ribas}, {Quirrenbach}, {Caballero}, {Amado}, {Carone}, {Khaimova}, {Kreidberg}, {Molaverdikhani}, {Montes}, {Morello}, {Nagel}, {Oshagh}, \& {Zechmeister}}]{Cont2021}
{Cont}, D., {Yan}, F., {Reiners}, A., {et~al.} 2021, \aap, 651, A33

\bibitem[{{Cont} {et~al.}(2022{\natexlab{a}}){Cont}, {Yan}, {Reiners}, {Nortmann}, {Molaverdikhani}, {Pall{\'e}}, {Henning}, {Ribas}, {Quirrenbach}, {Caballero}, {Amado}, {Czesla}, {Lesjak}, {L{\'o}pez-Puertas}, {Molli{\`e}re}, {Montes}, {Morello}, {Nagel}, {Pedraz}, \& {S{\'a}nchez-L{\'o}pez}}]{Cont2022b}
{Cont}, D., {Yan}, F., {Reiners}, A., {et~al.} 2022{\natexlab{a}}, \aap, 668, A53

\bibitem[{{Cont} {et~al.}(2022{\natexlab{b}}){Cont}, {Yan}, {Reiners}, {Nortmann}, {Molaverdikhani}, {Pall{\'e}}, {Stangret}, {Henning}, {Ribas}, {Quirrenbach}, {Caballero}, {Zapatero Osorio}, {Amado}, {Aceituno}, {Casasayas-Barris}, {Czesla}, {Kaminski}, {L{\'o}pez-Puertas}, {Montes}, {Morales}, {Morello}, {Nagel}, {S{\'a}nchez-L{\'o}pez}, {Sedaghati}, \& {Zechmeister}}]{Cont2022a}
{Cont}, D., {Yan}, F., {Reiners}, A., {et~al.} 2022{\natexlab{b}}, \aap, 657, L2

\bibitem[{{Cowan} \& {Agol}(2011)}]{CowanAgol2011}
{Cowan}, N.~B. \& {Agol}, E. 2011, \apj, 726, 82

\bibitem[{{Czesla} {et~al.}(2024){Czesla}, {Lamp{\'o}n}, {Cont}, {Lesjak}, {Orell-Miquel}, {Sanz-Forcada}, {Nagel}, {Nortmann}, {Molaverdikhani}, {L{\'o}pez-Puertas}, {Yan}, {Quirrenbach}, {Caballero}, {Pall{\'e}}, {Aceituno}, {Amado}, {Henning}, {Khalafinejad}, {Montes}, {Reiners}, {Ribas}, \& {Schweitzer}}]{Czesla2024}
{Czesla}, S., {Lamp{\'o}n}, M., {Cont}, D., {et~al.} 2024, \aap, 683, A67

\bibitem[{{Czesla} {et~al.}(2019){Czesla}, {Schr{\"o}ter}, {Schneider}, {Huber}, {Pfeifer}, {Andreasen}, \& {Zechmeister}}]{Czesla2019}
{Czesla}, S., {Schr{\"o}ter}, S., {Schneider}, C.~P., {et~al.} 2019, {PyA: Python astronomy-related packages}, Astrophysics Source Code Library, record ascl:1906.010

\bibitem[{{D{\'\i}az} {et~al.}(2011){D{\'\i}az}, {Gonz{\'a}lez}, {Levato}, \& {Grosso}}]{Diaz2011}
{D{\'\i}az}, C.~G., {Gonz{\'a}lez}, J.~F., {Levato}, H., \& {Grosso}, M. 2011, \aap, 531, A143

\bibitem[{{Dorn} {et~al.}(2023){Dorn}, {Bristow}, {Smoker}, {Rodler}, {Lavail}, {Accardo}, {van den Ancker}, {Baade}, {Baruffolo}, {Courtney-Barrer}, {Blanco}, {Brucalassi}, {Cumani}, {Follert}, {Haimerl}, {Hatzes}, {Haug}, {Heiter}, {Hinterschuster}, {Hubin}, {Ives}, {Jung}, {Jones}, {Kaeufl}, {Kirchbauer}, {Klein}, {Kochukhov}, {Korhonen}, {K{\"o}hler}, {Lizon}, {Moins}, {Molina-Conde}, {Marquart}, {Neeser}, {Oliva}, {Pallanca}, {Pasquini}, {Paufique}, {Piskunov}, {Reiners}, {Schneller}, {Schmutzer}, {Seemann}, {Slumstrup}, {Smette}, {Stegmeier}, {Stempels}, {Tordo}, {Valenti}, {Valenzuela}, {Vernet}, {Vinther}, \& {Wehrhahn}}]{Dorn2023}
{Dorn}, R.~J., {Bristow}, P., {Smoker}, J.~V., {et~al.} 2023, \aap, 671, A24

\bibitem[{{Finnerty} {et~al.}(2023){Finnerty}, {Schofield}, {Sappey}, {Xuan}, {Ruffio}, {Wang}, {Delorme}, {Blake}, {Buzard}, {Fitzgerald}, {Baker}, {Bartos}, {Bond}, {Calvin}, {Cetre}, {Doppmann}, {Echeverri}, {Jovanovic}, {Liberman}, {L{\'o}pez}, {Martin}, {Mawet}, {Morris}, {Pezzato}, {Phillips}, {Ragland}, {Skemer}, {Venenciano}, {Wallace}, {Wallack}, {Wang}, \& {Wizinowich}}]{Finnerty2023}
{Finnerty}, L., {Schofield}, T., {Sappey}, B., {et~al.} 2023, \aj, 166, 31

\bibitem[{Foreman-Mackey(2016)}]{ForemanMackey2016}
Foreman-Mackey, D. 2016, The Journal of Open Source Software, 1, 24

\bibitem[{{Foreman-Mackey} {et~al.}(2013){Foreman-Mackey}, {Hogg}, {Lang}, \& {Goodman}}]{Foreman-Mackey2013}
{Foreman-Mackey}, D., {Hogg}, D.~W., {Lang}, D., \& {Goodman}, J. 2013, \pasp, 125, 306

\bibitem[{{Fortney} {et~al.}(2021){Fortney}, {Dawson}, \& {Komacek}}]{Fortney2021}
{Fortney}, J.~J., {Dawson}, R.~I., \& {Komacek}, T.~D. 2021, Journal of Geophysical Research (Planets), 126, e06629

\bibitem[{{Fortney} {et~al.}(2008){Fortney}, {Lodders}, {Marley}, \& {Freedman}}]{Fortney2008}
{Fortney}, J.~J., {Lodders}, K., {Marley}, M.~S., \& {Freedman}, R.~S. 2008, \apj, 678, 1419

\bibitem[{{Fossati} {et~al.}(2023){Fossati}, {Biassoni}, {Cappello}, {Borsa}, {Shulyak}, {Bonomo}, {Gandolfi}, {Haardt}, {Koskinen}, {Lanza}, {Nascimbeni}, {Sicilia}, {Young}, {Aresu}, {Bignamini}, {Brogi}, {Carleo}, {Claudi}, {Cosentino}, {Guilluy}, {Knapic}, {Malavolta}, {Mancini}, {Nardiello}, {Pinamonti}, {Pino}, {Poretti}, {Rainer}, {Rigamonti}, \& {Sozzetti}}]{Fossati2023}
{Fossati}, L., {Biassoni}, F., {Cappello}, G.~M., {et~al.} 2023, \aap, 676, A99

\bibitem[{{Fossati} {et~al.}(2021){Fossati}, {Young}, {Shulyak}, {Koskinen}, {Huang}, {Cubillos}, {France}, \& {Sreejith}}]{Fossati2021}
{Fossati}, L., {Young}, M.~E., {Shulyak}, D., {et~al.} 2021, \aap, 653, A52

\bibitem[{{Gaia Collaboration} {et~al.}(2018){Gaia Collaboration}, {Brown}, {Vallenari}, {Prusti}, {de Bruijne}, {Babusiaux}, {Bailer-Jones}, {Biermann}, {Evans}, {Eyer}, {Jansen}, {Jordi}, {Klioner}, {Lammers}, {Lindegren}, {Luri}, {Mignard}, {Panem}, {Pourbaix}, {Randich}, {Sartoretti}, {Siddiqui}, {Soubiran}, {van Leeuwen}, {Walton}, {Arenou}, {Bastian}, {Cropper}, {Drimmel}, {Katz}, {Lattanzi}, {Bakker}, {Cacciari}, {Casta{\~n}eda}, {Chaoul}, {Cheek}, {De Angeli}, {Fabricius}, {Guerra}, {Holl}, {Masana}, {Messineo}, {Mowlavi}, {Nienartowicz}, {Panuzzo}, {Portell}, {Riello}, {Seabroke}, {Tanga}, {Th{\'e}venin}, {Gracia-Abril}, {Comoretto}, {Garcia-Reinaldos}, {Teyssier}, {Altmann}, {Andrae}, {Audard}, {Bellas-Velidis}, {Benson}, {Berthier}, {Blomme}, {Burgess}, {Busso}, {Carry}, {Cellino}, {Clementini}, {Clotet}, {Creevey}, {Davidson}, {De Ridder}, {Delchambre}, {Dell'Oro}, {Ducourant}, {Fern{\'a}ndez-Hern{\'a}ndez}, {Fouesneau}, {Fr{\'e}mat}, {Galluccio}, {Garc{\'\i}a-Torres},
  {Gonz{\'a}lez-N{\'u}{\~n}ez}, {Gonz{\'a}lez-Vidal}, {Gosset}, {Guy}, {Halbwachs}, {Hambly}, {Harrison}, {Hern{\'a}ndez}, {Hestroffer}, {Hodgkin}, {Hutton}, {Jasniewicz}, {Jean-Antoine-Piccolo}, {Jordan}, {Korn}, {Krone-Martins}, {Lanzafame}, {Lebzelter}, {L{\"o}ffler}, {Manteiga}, {Marrese}, {Mart{\'\i}n-Fleitas}, {Moitinho}, {Mora}, {Muinonen}, {Osinde}, {Pancino}, {Pauwels}, {Petit}, {Recio-Blanco}, {Richards}, {Rimoldini}, {Robin}, {Sarro}, {Siopis}, {Smith}, {Sozzetti}, {S{\"u}veges}, {Torra}, {van Reeven}, {Abbas}, {Abreu Aramburu}, {Accart}, {Aerts}, {Altavilla}, {{\'A}lvarez}, {Alvarez}, {Alves}, {Anderson}, {Andrei}, {Anglada Varela}, {Antiche}, {Antoja}, {Arcay}, {Astraatmadja}, {Bach}, {Baker}, {Balaguer-N{\'u}{\~n}ez}, {Balm}, {Barache}, {Barata}, {Barbato}, {Barblan}, {Barklem}, {Barrado}, {Barros}, {Barstow}, {Bartholom{\'e} Mu{\~n}oz}, {Bassilana}, {Becciani}, {Bellazzini}, {Berihuete}, {Bertone}, {Bianchi}, {Bienaym{\'e}}, {Blanco-Cuaresma}, {Boch}, {Boeche}, {Bombrun}, {Borrachero},
  {Bossini}, {Bouquillon}, {Bourda}, {Bragaglia}, {Bramante}, {Breddels}, {Bressan}, {Brouillet}, {Br{\"u}semeister}, {Brugaletta}, {Bucciarelli}, {Burlacu}, {Busonero}, {Butkevich}, {Buzzi}, {Caffau}, {Cancelliere}, {Cannizzaro}, {Cantat-Gaudin}, {Carballo}, {Carlucci}, {Carrasco}, {Casamiquela}, {Castellani}, {Castro-Ginard}, {Charlot}, {Chemin}, {Chiavassa}, {Cocozza}, {Costigan}, {Cowell}, {Crifo}, {Crosta}, {Crowley}, {Cuypers}, {Dafonte}, {Damerdji}, {Dapergolas}, {David}, {David}, {de Laverny}, \& {De Luise}}]{Gaia2018}
{Gaia Collaboration}, {Brown}, A.~G.~A., {Vallenari}, A., {et~al.} 2018, \aap, 616, A1

\bibitem[{{Gaia Collaboration} {et~al.}(2023){Gaia Collaboration}, {Vallenari}, {Brown}, {Prusti}, {de Bruijne}, {Arenou}, {Babusiaux}, {Biermann}, {Creevey}, {Ducourant}, {Evans}, {Eyer}, {Guerra}, {Hutton}, {Jordi}, {Klioner}, {Lammers}, {Lindegren}, {Luri}, {Mignard}, {Panem}, {Pourbaix}, {Randich}, {Sartoretti}, {Soubiran}, {Tanga}, {Walton}, {Bailer-Jones}, {Bastian}, {Drimmel}, {Jansen}, {Katz}, {Lattanzi}, {van Leeuwen}, {Bakker}, {Cacciari}, {Casta{\~n}eda}, {De Angeli}, {Fabricius}, {Fouesneau}, {Fr{\'e}mat}, {Galluccio}, {Guerrier}, {Heiter}, {Masana}, {Messineo}, {Mowlavi}, {Nicolas}, {Nienartowicz}, {Pailler}, {Panuzzo}, {Riclet}, {Roux}, {Seabroke}, {Sordo}, {Th{\'e}venin}, {Gracia-Abril}, {Portell}, {Teyssier}, {Altmann}, {Andrae}, {Audard}, {Bellas-Velidis}, {Benson}, {Berthier}, {Blomme}, {Burgess}, {Busonero}, {Busso}, {C{\'a}novas}, {Carry}, {Cellino}, {Cheek}, {Clementini}, {Damerdji}, {Davidson}, {de Teodoro}, {Nu{\~n}ez Campos}, {Delchambre}, {Dell'Oro}, {Esquej},
  {Fern{\'a}ndez-Hern{\'a}ndez}, {Fraile}, {Garabato}, {Garc{\'\i}a-Lario}, {Gosset}, {Haigron}, {Halbwachs}, {Hambly}, {Harrison}, {Hern{\'a}ndez}, {Hestroffer}, {Hodgkin}, {Holl}, {Jan{\ss}en}, {Jevardat de Fombelle}, {Jordan}, {Krone-Martins}, {Lanzafame}, {L{\"o}ffler}, {Marchal}, {Marrese}, {Moitinho}, {Muinonen}, {Osborne}, {Pancino}, {Pauwels}, {Recio-Blanco}, {Reyl{\'e}}, {Riello}, {Rimoldini}, {Roegiers}, {Rybizki}, {Sarro}, {Siopis}, {Smith}, {Sozzetti}, {Utrilla}, {van Leeuwen}, {Abbas}, {{\'A}brah{\'a}m}, {Abreu Aramburu}, {Aerts}, {Aguado}, {Ajaj}, {Aldea-Montero}, {Altavilla}, {{\'A}lvarez}, {Alves}, {Anders}, {Anderson}, {Anglada Varela}, {Antoja}, {Baines}, {Baker}, {Balaguer-N{\'u}{\~n}ez}, {Balbinot}, {Balog}, {Barache}, {Barbato}, {Barros}, {Barstow}, {Bartolom{\'e}}, {Bassilana}, {Bauchet}, {Becciani}, {Bellazzini}, {Berihuete}, {Bernet}, {Bertone}, {Bianchi}, {Binnenfeld}, {Blanco-Cuaresma}, {Blazere}, {Boch}, {Bombrun}, {Bossini}, {Bouquillon}, {Bragaglia}, {Bramante}, {Breedt},
  {Bressan}, {Brouillet}, {Brugaletta}, {Bucciarelli}, {Burlacu}, {Butkevich}, {Buzzi}, {Caffau}, {Cancelliere}, {Cantat-Gaudin}, {Carballo}, {Carlucci}, {Carnerero}, {Carrasco}, {Casamiquela}, {Castellani}, {Castro-Ginard}, {Chaoul}, {Charlot}, {Chemin}, {Chiaramida}, {Chiavassa}, {Chornay}, {Comoretto}, {Contursi}, {Cooper}, {Cornez}, {Cowell}, {Crifo}, {Cropper}, {Crosta}, {Crowley}, {Dafonte}, {Dapergolas}, {David}, {David}, {de Laverny}, {De Luise}, \& {De March}}]{Gaia2023}
{Gaia Collaboration}, {Vallenari}, A., {Brown}, A.~G.~A., {et~al.} 2023, \aap, 674, A1

\bibitem[{{Gandhi} {et~al.}(2023){Gandhi}, {de Regt}, {Snellen}, {Zhang}, {Rugers}, {van Leur}, \& {Bosschaart}}]{Gandhi2023}
{Gandhi}, S., {de Regt}, S., {Snellen}, I., {et~al.} 2023, \apjl, 957, L36

\bibitem[{{Gaudi} {et~al.}(2017){Gaudi}, {Stassun}, {Collins}, {Beatty}, {Zhou}, {Latham}, {Bieryla}, {Eastman}, {Siverd}, {Crepp}, {Gonzales}, {Stevens}, {Buchhave}, {Pepper}, {Johnson}, {Colon}, {Jensen}, {Rodriguez}, {Bozza}, {Novati}, {D'Ago}, {Dumont}, {Ellis}, {Gaillard}, {Jang-Condell}, {Kasper}, {Fukui}, {Gregorio}, {Ito}, {Kielkopf}, {Manner}, {Matt}, {Narita}, {Oberst}, {Reed}, {Scarpetta}, {Stephens}, {Yeigh}, {Zambelli}, {Fulton}, {Howard}, {James}, {Penny}, {Bayliss}, {Curtis}, {Depoy}, {Esquerdo}, {Gould}, {Joner}, {Kuhn}, {Labadie-Bartz}, {Lund}, {Marshall}, {McLeod}, {Pogge}, {Relles}, {Stockdale}, {Tan}, {Trueblood}, \& {Trueblood}}]{Gaudi2017}
{Gaudi}, B.~S., {Stassun}, K.~G., {Collins}, K.~A., {et~al.} 2017, \nat, 546, 514

\bibitem[{{Gibson} {et~al.}(2020){Gibson}, {Merritt}, {Nugroho}, {Cubillos}, {de Mooij}, {Mikal-Evans}, {Fossati}, {Lothringer}, {Nikolov}, {Sing}, {Spake}, {Watson}, \& {Wilson}}]{Gibson2020}
{Gibson}, N.~P., {Merritt}, S., {Nugroho}, S.~K., {et~al.} 2020, \mnras, 493, 2215

\bibitem[{{Gibson} {et~al.}(2022){Gibson}, {Nugroho}, {Lothringer}, {Maguire}, \& {Sing}}]{Gibson2022}
{Gibson}, N.~P., {Nugroho}, S.~K., {Lothringer}, J., {Maguire}, C., \& {Sing}, D.~K. 2022, \mnras, 512, 4618

\bibitem[{{Grimm} \& {Heng}(2015)}]{GrimmHeng2015}
{Grimm}, S.~L. \& {Heng}, K. 2015, \apj, 808, 182

\bibitem[{{Guo} {et~al.}(2024){Guo}, {Yan}, {Nortmann}, {Cont}, {Reiners}, {Pall{\'e}}, {Shulyak}, {Molaverdikhani}, {Henning}, {Chen}, {Stangret}, {Czesla}, {Lesjak}, {L{\'o}pez-Puertas}, {Ribas}, {Quirrenbach}, {Caballero}, {Amado}, {Blazek}, {Montes}, {Morales}, {Nagel}, \& {Zapatero Osorio}}]{Guo2024}
{Guo}, B., {Yan}, F., {Nortmann}, L., {et~al.} 2024, \aap, 687, A103

\bibitem[{{Harre} {et~al.}(2024){Harre}, {Smith}, {Barros}, {Singh}, {Korth}, {Brandeker}, {Collier Cameron}, {Lendl}, {Wilson}, {Borsato}, {Csizmadia}, {Cabrera}, {Parviainen}, {Correia}, {Akinsanmi}, {Rosario}, {Leonardi}, {Serrano}, {Alibert}, {Alonso}, {Asquier}, {B{\'a}rczy}, {Barrado Navascues}, {Baumjohann}, {Benz}, {Billot}, {Broeg}, {Busch}, {Cubillos}, {Davies}, {Deleuil}, {Deline}, {Delrez}, {Demangeon}, {Demory}, {Derekas}, {Edwards}, {Ehrenreich}, {Erikson}, {Fortier}, {Fossati}, {Fridlund}, {Gandolfi}, {Gazeas}, {Gillon}, {G{\"u}del}, {G{\"u}nther}, {Heitzmann}, {Helling}, {Isaak}, {Kiss}, {Lam}, {Laskar}, {Lecavelier des Etangs}, {Magrin}, {Maxted}, {Mer{\'\i}n}, {Mordasini}, {Nascimbeni}, {Olofsson}, {Ottensamer}, {Pagano}, {Pall{\'e}}, {Peter}, {Piazza}, {Piotto}, {Pollacco}, {Queloz}, {Ragazzoni}, {Rando}, {Rauer}, {Ribas}, {Santos}, {Scandariato}, {S{\'e}gransan}, {Simon}, {Sousa}, {Stalport}, {Sulis}, {Szab{\'o}}, {Udry}, {Ulmer}, {Van Grootel}, {Venturini}, {Villaver}, {Viotto}, {Walton},
  {West}, \& {Westerdorff}}]{Harre2024}
{Harre}, J.~V., {Smith}, A.~M.~S., {Barros}, S.~C.~C., {et~al.} 2024, \aap, 692, A254

\bibitem[{{Harris} {et~al.}(2020){Harris}, {Millman}, {van der Walt}, {Gommers}, {Virtanen}, {Cournapeau}, {Wieser}, {Taylor}, {Berg}, {Smith}, {Kern}, {Picus}, {Hoyer}, {van Kerkwijk}, {Brett}, {Haldane}, {del R{\'\i}o}, {Wiebe}, {Peterson}, {G{\'e}rard-Marchant}, {Sheppard}, {Reddy}, {Weckesser}, {Abbasi}, {Gohlke}, \& {Oliphant}}]{Harris2020}
{Harris}, C.~R., {Millman}, K.~J., {van der Walt}, S.~J., {et~al.} 2020, \nat, 585, 357

\bibitem[{{Hoeijmakers} {et~al.}(2024){Hoeijmakers}, {Kitzmann}, {Morris}, {Prinoth}, {Borsato}, {Thorsbro}, {Pino}, {Lee}, {Ak{\i}n}, {Seidel}, {Birkby}, {Allart}, \& {Heng}}]{Hoeijmakers2024}
{Hoeijmakers}, H.~J., {Kitzmann}, D., {Morris}, B.~M., {et~al.} 2024, \aap, 685, A139

\bibitem[{{Hubeny} {et~al.}(2003){Hubeny}, {Burrows}, \& {Sudarsky}}]{Hubeny2003}
{Hubeny}, I., {Burrows}, A., \& {Sudarsky}, D. 2003, \apj, 594, 1011

\bibitem[{{Hunter}(2007)}]{Hunter2007}
{Hunter}, J.~D. 2007, Computing in Science and Engineering, 9, 90

\bibitem[{{Hut}(1981)}]{Hut1981}
{Hut}, P. 1981, \aap, 99, 126

\bibitem[{{Johnson} {et~al.}(2023){Johnson}, {Wang}, {Asnodkar}, {Bonomo}, {Gaudi}, {Henning}, {Ilyin}, {Keles}, {Malavolta}, {Mallonn}, {Molaverdikhani}, {Nascimbeni}, {Patience}, {Poppenhaeger}, {Scandariato}, {Schlawin}, {Shkolnik}, {Sicilia}, {Sozzetti}, {Strassmeier}, {Veillet}, \& {Yan}}]{Johnson2023}
{Johnson}, M.~C., {Wang}, J., {Asnodkar}, A.~P., {et~al.} 2023, \aj, 165, 157

\bibitem[{{Kasper} {et~al.}(2021){Kasper}, {Bean}, {Line}, {Seifahrt}, {St{\"u}rmer}, {Pino}, {D{\'e}sert}, \& {Brogi}}]{Kasper2021}
{Kasper}, D., {Bean}, J.~L., {Line}, M.~R., {et~al.} 2021, \apjl, 921, L18

\bibitem[{{Keating} {et~al.}(2019){Keating}, {Cowan}, \& {Dang}}]{Keating2019}
{Keating}, D., {Cowan}, N.~B., \& {Dang}, L. 2019, Nature Astronomy, 3, 1092

\bibitem[{{Kitzmann} {et~al.}(2018){Kitzmann}, {Heng}, {Rimmer}, {Hoeijmakers}, {Tsai}, {Malik}, {Lendl}, {Deitrick}, \& {Demory}}]{Kitzmann2018}
{Kitzmann}, D., {Heng}, K., {Rimmer}, P.~B., {et~al.} 2018, \apj, 863, 183

\bibitem[{{Komacek} \& {Showman}(2016)}]{KomacekShowman2016}
{Komacek}, T.~D. \& {Showman}, A.~P. 2016, \apj, 821, 16

\bibitem[{{Komacek} \& {Tan}(2018)}]{KomacekTan2018}
{Komacek}, T.~D. \& {Tan}, X. 2018, Research Notes of the American Astronomical Society, 2, 36

\bibitem[{{Kreidberg} {et~al.}(2018){Kreidberg}, {Line}, {Parmentier}, {Stevenson}, {Louden}, {Bonnefoy}, {Faherty}, {Henry}, {Williamson}, {Stassun}, {Beatty}, {Bean}, {Fortney}, {Showman}, {D{\'e}sert}, \& {Arcangeli}}]{Kreidberg2018}
{Kreidberg}, L., {Line}, M.~R., {Parmentier}, V., {et~al.} 2018, \aj, 156, 17

\bibitem[{{Kurucz}(2018)}]{Kurucz2018}
{Kurucz}, R.~L. 2018, in Astronomical Society of the Pacific Conference Series, Vol. 515, Workshop on Astrophysical Opacities, 47

\bibitem[{{Landman} {et~al.}(2021){Landman}, {S{\'a}nchez-L{\'o}pez}, {Molli{\`e}re}, {Kesseli}, {Louca}, \& {Snellen}}]{Landman2021}
{Landman}, R., {S{\'a}nchez-L{\'o}pez}, A., {Molli{\`e}re}, P., {et~al.} 2021, \aap, 656, A119

\bibitem[{Lavail(2025)}]{Lavail2025}
Lavail, A. 2025, CRIRES+ and CARMENES reduced spectroscopic observations of TOI-2109

\bibitem[{{Lesjak} {et~al.}(2025){Lesjak}, {Nortmann}, {Cont}, {Yan}, {Reiners}, {Piskunov}, {Hatzes}, {Boldt-Christmas}, {Czesla}, {Lavail}, {Nagel}, {Rains}, {Rengel}, {Seemann}, \& {Shulyak}}]{Lesjak2025}
{Lesjak}, F., {Nortmann}, L., {Cont}, D., {et~al.} 2025, \aap, 693, A72

\bibitem[{{Li} {et~al.}(2015){Li}, {Gordon}, {Rothman}, {Tan}, {Hu}, {Kassi}, {Campargue}, \& {Medvedev}}]{Li2015}
{Li}, G., {Gordon}, I.~E., {Rothman}, L.~S., {et~al.} 2015, \apjs, 216, 15

\bibitem[{{Line} {et~al.}(2021){Line}, {Brogi}, {Bean}, {Gandhi}, {Zalesky}, {Parmentier}, {Smith}, {Mace}, {Mansfield}, {Kempton}, {Fortney}, {Shkolnik}, {Patience}, {Rauscher}, {D{\'e}sert}, \& {Wardenier}}]{Line2021}
{Line}, M.~R., {Brogi}, M., {Bean}, J.~L., {et~al.} 2021, \nat, 598, 580

\bibitem[{{Lothringer} {et~al.}(2018){Lothringer}, {Barman}, \& {Koskinen}}]{Lothringer2018}
{Lothringer}, J.~D., {Barman}, T., \& {Koskinen}, T. 2018, \apj, 866, 27

\bibitem[{{Malik} {et~al.}(2017){Malik}, {Grosheintz}, {Mendon{\c{c}}a}, {Grimm}, {Lavie}, {Kitzmann}, {Tsai}, {Burrows}, {Kreidberg}, {Bedell}, {Bean}, {Stevenson}, \& {Heng}}]{Malik2017}
{Malik}, M., {Grosheintz}, L., {Mendon{\c{c}}a}, J.~M., {et~al.} 2017, \aj, 153, 56

\bibitem[{{Malik} {et~al.}(2019){Malik}, {Kitzmann}, {Mendon{\c{c}}a}, {Grimm}, {Marleau}, {Linder}, {Tsai}, \& {Heng}}]{Malik2019}
{Malik}, M., {Kitzmann}, D., {Mendon{\c{c}}a}, J.~M., {et~al.} 2019, \aj, 157, 170

\bibitem[{{Matthews} {et~al.}(2024){Matthews}, {Watson}, {de Mooij}, {Marsh}, {Brogi}, {Merritt}, {Smith}, \& {Steeghs}}]{Matthews2024}
{Matthews}, S.~M., {Watson}, C.~A., {de Mooij}, E.~J.~W., {et~al.} 2024, \mnras, 531, 3800

\bibitem[{{Miller-Ricci Kempton} \& {Rauscher}(2012)}]{MillerRicciKempton2012}
{Miller-Ricci Kempton}, E. \& {Rauscher}, E. 2012, \apj, 751, 117

\bibitem[{{Molaverdikhani} {et~al.}(2020){Molaverdikhani}, {Helling}, {Lew}, {MacDonald}, {Samra}, {Iro}, {Woitke}, \& {Parmentier}}]{Molaverdikhani2020}
{Molaverdikhani}, K., {Helling}, C., {Lew}, B.~W.~P., {et~al.} 2020, \aap, 635, A31

\bibitem[{{Molli{\`e}re} {et~al.}(2019){Molli{\`e}re}, {Wardenier}, {van Boekel}, {Henning}, {Molaverdikhani}, \& {Snellen}}]{Molliere2019}
{Molli{\`e}re}, P., {Wardenier}, J.~P., {van Boekel}, R., {et~al.} 2019, \aap, 627, A67

\bibitem[{{Morello} {et~al.}(2023){Morello}, {Changeat}, {Dyrek}, {Lagage}, \& {Tan}}]{Morello2023}
{Morello}, G., {Changeat}, Q., {Dyrek}, A., {Lagage}, P.~O., \& {Tan}, J.~C. 2023, \aap, 676, A54

\bibitem[{{Morello} {et~al.}(2019){Morello}, {Danielski}, {Dickens}, {Tremblin}, \& {Lagage}}]{Morello2019}
{Morello}, G., {Danielski}, C., {Dickens}, D., {Tremblin}, P., \& {Lagage}, P.~O. 2019, \aj, 157, 205

\bibitem[{{Mraz} {et~al.}(2024){Mraz}, {Darveau-Bernier}, {Boucher}, {Cowan}, {Lafreni{\`e}re}, \& {Cadieux}}]{Mraz2024}
{Mraz}, G., {Darveau-Bernier}, A., {Boucher}, A., {et~al.} 2024, \apjl, 975, L42

\bibitem[{{Nugroho} {et~al.}(2021){Nugroho}, {Kawahara}, {Gibson}, {de Mooij}, {Hirano}, {Kotani}, {Kawashima}, {Masuda}, {Brogi}, {Birkby}, {Watson}, {Tamura}, {Zwintz}, {Harakawa}, {Kudo}, {Kuzuhara}, {Hodapp}, {Ishizuka}, {Jacobson}, {Konishi}, {Kurokawa}, {Nishikawa}, {Omiya}, {Serizawa}, {Ueda}, \& {Vievard}}]{Nugroho2021}
{Nugroho}, S.~K., {Kawahara}, H., {Gibson}, N.~P., {et~al.} 2021, \apjl, 910, L9

\bibitem[{{Nugroho} {et~al.}(2017){Nugroho}, {Kawahara}, {Masuda}, {Hirano}, {Kotani}, \& {Tajitsu}}]{Nugroho2017}
{Nugroho}, S.~K., {Kawahara}, H., {Masuda}, K., {et~al.} 2017, \aj, 154, 221

\bibitem[{{Parmentier} {et~al.}(2018){Parmentier}, {Line}, {Bean}, {Mansfield}, {Kreidberg}, {Lupu}, {Visscher}, {D{\'e}sert}, {Fortney}, {Deleuil}, {Arcangeli}, {Showman}, \& {Marley}}]{Parmentier2018}
{Parmentier}, V., {Line}, M.~R., {Bean}, J.~L., {et~al.} 2018, \aap, 617, A110

\bibitem[{{Parmentier} {et~al.}(2013){Parmentier}, {Showman}, \& {Lian}}]{Parmentier2013}
{Parmentier}, V., {Showman}, A.~P., \& {Lian}, Y. 2013, \aap, 558, A91

\bibitem[{{Pelletier} {et~al.}(2023){Pelletier}, {Benneke}, {Ali-Dib}, {Prinoth}, {Kasper}, {Seifahrt}, {Bean}, {Debras}, {Klein}, {Bazinet}, {Hoeijmakers}, {Kesseli}, {Lim}, {Carmona}, {Pino}, {Casasayas-Barris}, {Hood}, \& {St{\"u}rmer}}]{Pelletier2023}
{Pelletier}, S., {Benneke}, B., {Ali-Dib}, M., {et~al.} 2023, \nat, 619, 491

\bibitem[{{Pelletier} {et~al.}(2025){Pelletier}, {Benneke}, {Chachan}, {Bazinet}, {Allart}, {Hoeijmakers}, {Lavail}, {Prinoth}, {Coulombe}, {Lothringer}, {Parmentier}, {Smith}, {Borsato}, \& {Thorsbro}}]{Pelletier2025}
{Pelletier}, S., {Benneke}, B., {Chachan}, Y., {et~al.} 2025, \aj, 169, 10

\bibitem[{{Pelletier} {et~al.}(2021){Pelletier}, {Benneke}, {Darveau-Bernier}, {Boucher}, {Cook}, {Piaulet}, {Coulombe}, {Artigau}, {Lafreni{\`e}re}, {Delisle}, {Allart}, {Doyon}, {Donati}, {Fouqu{\'e}}, {Moutou}, {Cadieux}, {Delfosse}, {H{\'e}brard}, {Martins}, {Martioli}, \& {Vandal}}]{Pelletier2021}
{Pelletier}, S., {Benneke}, B., {Darveau-Bernier}, A., {et~al.} 2021, \aj, 162, 73

\bibitem[{{Pino} {et~al.}(2020){Pino}, {D{\'e}sert}, {Brogi}, {Malavolta}, {Wyttenbach}, {Line}, {Hoeijmakers}, {Fossati}, {Bonomo}, {Nascimbeni}, {Panwar}, {Affer}, {Benatti}, {Biazzo}, {Bignamini}, {Borsa}, {Carleo}, {Claudi}, {Cosentino}, {Covino}, {Damasso}, {Desidera}, {Giacobbe}, {Harutyunyan}, {Lanza}, {Leto}, {Maggio}, {Maldonado}, {Mancini}, {Micela}, {Molinari}, {Pagano}, {Piotto}, {Poretti}, {Rainer}, {Scandariato}, {Sozzetti}, {Allart}, {Borsato}, {Bruno}, {Di Fabrizio}, {Ehrenreich}, {Fiorenzano}, {Frustagli}, {Lavie}, {Lovis}, {Magazz{\`u}}, {Nardiello}, {Pedani}, \& {Smareglia}}]{Pino2020}
{Pino}, L., {D{\'e}sert}, J.-M., {Brogi}, M., {et~al.} 2020, \apjl, 894, L27

\bibitem[{{Polyansky} {et~al.}(2018){Polyansky}, {Kyuberis}, {Zobov}, {Tennyson}, {Yurchenko}, \& {Lodi}}]{Polyansky2018}
{Polyansky}, O.~L., {Kyuberis}, A.~A., {Zobov}, N.~F., {et~al.} 2018, \mnras, 480, 2597

\bibitem[{{Prinoth} {et~al.}(2022){Prinoth}, {Hoeijmakers}, {Kitzmann}, {Sandvik}, {Seidel}, {Lendl}, {Borsato}, {Thorsbro}, {Anderson}, {Barrado}, {Kravchenko}, {Allart}, {Bourrier}, {Cegla}, {Ehrenreich}, {Fisher}, {Lovis}, {Guzm{\'a}n-Mesa}, {Grimm}, {Hooton}, {Morris}, {Oreshenko}, {Pino}, \& {Heng}}]{Prinoth2022}
{Prinoth}, B., {Hoeijmakers}, H.~J., {Kitzmann}, D., {et~al.} 2022, Nature Astronomy, 6, 449

\bibitem[{{Quirrenbach} {et~al.}(2018){Quirrenbach}, {Amado}, {Ribas}, {Reiners}, {Caballero}, {Seifert}, {Aceituno}, {Azzaro}, {Baroch}, {Barrado}, {Bauer}, {Becerril}, {B{\`e}jar}, {Ben{\'\i}tez}, {Brinkm{\"o}ller}, {Cardona Guill{\'e}n}, {Cifuentes}, {Colom{\'e}}, {Cort{\'e}s-Contreras}, {Czesla}, {Dreizler}, {Fr{\"o}lich}, {Fuhrmeister}, {Galad{\'\i}-Enr{\'\i}quez}, {Gonz{\'a}lez Hern{\'a}ndez}, {Gonz{\'a}lez Peinado}, {Guenther}, {de Guindos}, {Hagen}, {Hatzes}, {Hauschildt}, {Helmling}, {Henning}, {Herbort}, {Hern{\'a}ndez Casta{\~n}o}, {Herrero}, {Hintz}, {Jeffers}, {Johnson}, {de Juan}, {Kaminski}, {Klahr}, {K{\"u}rster}, {Lafarga}, {Sairam}, {Lamp{\'o}n}, {Lara}, {Launhardt}, {L{\'o}pez del Fresno}, {L{\'o}pez-Puertas}, {Luque}, {Mandel}, {Marfil}, {Mart{\'\i}n}, {Mart{\'\i}n-Ruiz}, {Mathar}, {Montes}, {Morales}, {Nagel}, {Nortmann}, {Nowak}, {Pall{\'e}}, {Passegger}, {Pavlov}, {Pedraz}, {P{\'e}rez-Medialdea}, {Perger}, {Rebolo}, {Reffert}, {Rodr{\'\i}guez}, {Rodr{\'\i}guez L{\'o}pez}, {Rosich},
  {Sabotta}, {Sadegi}, {Salz}, {S{\'a}nchez-L{\'o}pez}, {Sanz-Forcada}, {Sarkis}, {Sch{\"a}fer}, {Schiller}, {Schmitt}, {Sch{\"o}fer}, {Schweitzer}, {Shulyak}, {Solano}, {Stahl}, {Tala Pinto}, {Trifonov}, {Zapatero Osorio}, {Yan}, {Zechmeister}, {Abell{\'a}n}, {Abril}, {Alonso-Floriano}, {Ammler-von Eiff}, {Anglada-Escud{\'e}}, {Anwand-Heerwart}, {Arroyo-Torres}, {Berdi{\~n}as}, {Bergondy}, {Bl{\"u}mcke}, {del Burgo}, {Cano}, {Carro}, {C{\'a}rdenas}, {Casal}, {Claret}, {D{\'\i}ez-Alonso}, {Doellinger}, {Dorda}, {Feiz}, {Fern{\'a}ndez}, {Ferro}, {Gaisn{\'e}}, {Gallardo}, {G{\'a}lvez-Ortiz}, {Garc{\'\i}a-Piquer}, {Garc{\'\i}a-Vargas}, {Garrido}, {Gesa}, {G{\'o}mez Galera}, {Gonz{\'a}lez-{\'A}lvarez}, {Gonz{\'a}lez-Cuesta}, {Grohnert}, {Gr{\"o}zinger}, {Gu{\`a}rdia}, {Guijarro}, {Hedrosa}, {Hermann}, {Hermelo}, {Hern{\'a}ndez Arab{\'\i}}, {Hern{\'a}ndez Hernando}, {Hidalgo}, {Holgado}, {Huber}, {Huber}, {Huke}, {Kehr}, {Kim}, {Klein}, {Kl{\"u}ter}, {Klutsch}, {Labarga}, {Labiche}, {Lamert}, {Laun}, {L{\'a}zaro},
  {Lemke}, {Lenzen}, {Llamas}, {Lizon}, {Lodieu}, {L{\'o}pez Gonz{\'a}lez}, {L{\'o}pez-Morales}, {L{\'o}pez Salas}, {L{\'o}pez-Santiago}, {Mag{\'a}n Madinabeitia}, {Mall}, {Mancini}, {Mar{\'\i}n Molina}, {Mart{\'\i}nez-Rodr{\'\i}guez}, {Maroto Fern{\'a}ndez}, {Marvin}, {Mirabet}, {Moreno-Raya}, {Moya}, {Mundt}, {Naranjo}, {Panduro}, {Pascual}, {P{\'e}rez-Calpena}, {Perryman}, {Pluto}, {Ram{\'o}n}, {Redondo}, {Reinhart}, {Rhode}, {Rix}, {Rodler}, {Rohloff}, {S{\'a}nchez-Blanco}, {S{\'a}nchez Carrasco}, {Sarmiento}, {Schmidt}, {Storz}, {Strachan}, {St{\"u}rmer}, {Su{\'a}rez}, {Tabernero}, {Tal-Or}, {Tulloch}, {Ulbrich}, {Veredas}, {Vico Linares}, {Vidal-Dasilva}, {Vilardell}, {Wagner}, {Winkler}, {Wolthoff}, \& {Xu}}]{Quirrenbach2018}
{Quirrenbach}, A., {Amado}, P.~J., {Ribas}, I., {et~al.} 2018, in Society of Photo-Optical Instrumentation Engineers (SPIE) Conference Series, Vol. 10702, Ground-based and Airborne Instrumentation for Astronomy VII, ed. C.~J. {Evans}, L.~{Simard}, \& H.~{Takami}, 107020W

\bibitem[{{Ramkumar} {et~al.}(2025){Ramkumar}, {Gibson}, {Nugroho}, {Fortune}, \& {Maguire}}]{Ramkumar2025}
{Ramkumar}, S., {Gibson}, N.~P., {Nugroho}, S.~K., {Fortune}, M., \& {Maguire}, C. 2025, \aap, 695, A110

\bibitem[{{Ramkumar} {et~al.}(2023){Ramkumar}, {Gibson}, {Nugroho}, {Maguire}, \& {Fortune}}]{Ramkumar2023}
{Ramkumar}, S., {Gibson}, N.~P., {Nugroho}, S.~K., {Maguire}, C., \& {Fortune}, M. 2023, \mnras, 525, 2985

\bibitem[{{Ridden-Harper} {et~al.}(2023){Ridden-Harper}, {de Mooij}, {Jayawardhana}, {Gibson}, {Karjalainen}, \& {Karjalainen}}]{Ridden-Harper2023}
{Ridden-Harper}, A., {de Mooij}, E., {Jayawardhana}, R., {et~al.} 2023, \aj, 165, 211

\bibitem[{{Roth} {et~al.}(2021){Roth}, {Drummond}, {H{\'e}brard}, {Tremblin}, {Goyal}, \& {Mayne}}]{Roth2021}
{Roth}, A., {Drummond}, B., {H{\'e}brard}, E., {et~al.} 2021, \mnras, 505, 4515

\bibitem[{{Schwartz} {et~al.}(2017){Schwartz}, {Kashner}, {Jovmir}, \& {Cowan}}]{Schwartz2017}
{Schwartz}, J.~C., {Kashner}, Z., {Jovmir}, D., \& {Cowan}, N.~B. 2017, \apj, 850, 154

\bibitem[{{Schwarz} {et~al.}(2015){Schwarz}, {Brogi}, {de Kok}, {Birkby}, \& {Snellen}}]{Schwarz2015}
{Schwarz}, H., {Brogi}, M., {de Kok}, R., {Birkby}, J., \& {Snellen}, I. 2015, \aap, 576, A111

\bibitem[{{Seidel} {et~al.}(2025){Seidel}, {Prinoth}, {Pino}, {dos Santos}, {Chakraborty}, {Parmentier}, {Sedaghati}, {Wardenier}, {Farret Jentink}, {Zapatero Osorio}, {Allart}, {Ehrenreich}, {Lendl}, {Roccetti}, {Damasceno}, {Bourrier}, {Lillo-Box}, {Hoeijmakers}, {Pall{\'e}}, {Santos}, {Su{\'a}rez Mascare{\~n}o}, {Sousa}, {Tabernero}, \& {Pepe}}]{Seidel2025}
{Seidel}, J.~V., {Prinoth}, B., {Pino}, L., {et~al.} 2025, \nat, 639, 902

\bibitem[{{Showman} {et~al.}(2013){Showman}, {Fortney}, {Lewis}, \& {Shabram}}]{Showman2013}
{Showman}, A.~P., {Fortney}, J.~J., {Lewis}, N.~K., \& {Shabram}, M. 2013, \apj, 762, 24

\bibitem[{{Shporer} {et~al.}(2019){Shporer}, {Wong}, {Huang}, {Line}, {Stassun}, {Fetherolf}, {Kane}, {Bouma}, {Daylan}, {G{\"u}enther}, {Ricker}, {Latham}, {Vanderspek}, {Seager}, {Winn}, {Jenkins}, {Glidden}, {Berta-Thompson}, {Ting}, {Li}, \& {Haworth}}]{Shporer2019}
{Shporer}, A., {Wong}, I., {Huang}, C.~X., {et~al.} 2019, \aj, 157, 178

\bibitem[{{Smette} {et~al.}(2015){Smette}, {Sana}, {Noll}, {Horst}, {Kausch}, {Kimeswenger}, {Barden}, {Szyszka}, {Jones}, {Gallenne}, {Vinther}, {Ballester}, \& {Taylor}}]{Smette2015}
{Smette}, A., {Sana}, H., {Noll}, S., {et~al.} 2015, \aap, 576, A77

\bibitem[{{Snellen} {et~al.}(2010){Snellen}, {de Kok}, {de Mooij}, \& {Albrecht}}]{Snellen2010}
{Snellen}, I. A.~G., {de Kok}, R.~J., {de Mooij}, E. J.~W., \& {Albrecht}, S. 2010, \nat, 465, 1049

\bibitem[{{Spiegel} {et~al.}(2009){Spiegel}, {Silverio}, \& {Burrows}}]{Spiegel2009}
{Spiegel}, D.~S., {Silverio}, K., \& {Burrows}, A. 2009, \apj, 699, 1487

\bibitem[{{Stock} {et~al.}(2022){Stock}, {Kitzmann}, \& {Patzer}}]{Stock2022}
{Stock}, J.~W., {Kitzmann}, D., \& {Patzer}, A. B.~C. 2022, \mnras, 517, 4070

\bibitem[{{Tamuz} {et~al.}(2005){Tamuz}, {Mazeh}, \& {Zucker}}]{Tamuz2005}
{Tamuz}, O., {Mazeh}, T., \& {Zucker}, S. 2005, \mnras, 356, 1466

\bibitem[{{Tan} \& {Komacek}(2019)}]{TanKomacek2019}
{Tan}, X. \& {Komacek}, T.~D. 2019, \apj, 886, 26

\bibitem[{{Tange}(2021)}]{Tange2021}
{Tange}, O. 2021, {GNU Parallel 20210822 ('Kabul')}, Zenodo

\bibitem[{{van der Velden}(2020)}]{vanderVelden2020}
{van der Velden}, E. 2020, The Journal of Open Source Software, 5, 2004

\bibitem[{{Virtanen} {et~al.}(2020){Virtanen}, {Gommers}, {Oliphant}, {Haberland}, {Reddy}, {Cournapeau}, {Burovski}, {Peterson}, {Weckesser}, {Bright}, {van der Walt}, {Brett}, {Wilson}, {Millman}, {Mayorov}, {Nelson}, {Jones}, {Kern}, {Larson}, {Carey}, {Polat}, {Feng}, {Moore}, {VanderPlas}, {Laxalde}, {Perktold}, {Cimrman}, {Henriksen}, {Quintero}, {Harris}, {Archibald}, {Ribeiro}, {Pedregosa}, {van Mulbregt}, \& {SciPy 1. 0 Contributors}}]{Virtanen2020}
{Virtanen}, P., {Gommers}, R., {Oliphant}, T.~E., {et~al.} 2020, Nature Methods, 17, 261

\bibitem[{{Wardenier} {et~al.}(2025){Wardenier}, {Parmentier}, {Lee}, \& {Line}}]{Wardenier2025}
{Wardenier}, J.~P., {Parmentier}, V., {Lee}, E. K.~H., \& {Line}, M.~R. 2025, arXiv e-prints, arXiv:2502.01606

\bibitem[{{Wong} {et~al.}(2021){Wong}, {Shporer}, {Zhou}, {Kitzmann}, {Komacek}, {Tan}, {Tronsgaard}, {Buchhave}, {Vissapragada}, {Greklek-McKeon}, {Rodriguez}, {Ahlers}, {Quinn}, {Furlan}, {Howell}, {Bieryla}, {Heng}, {Knutson}, {Collins}, {McLeod}, {Berlind}, {Brown}, {Calkins}, {de Leon}, {Esparza-Borges}, {Esquerdo}, {Fukui}, {Gan}, {Girardin}, {Gnilka}, {Ikoma}, {Jensen}, {Kielkopf}, {Kodama}, {Kurita}, {Lester}, {Lewin}, {Marino}, {Murgas}, {Narita}, {Pall{\'e}}, {Schwarz}, {Stassun}, {Tamura}, {Watanabe}, {Benneke}, {Ricker}, {Latham}, {Vanderspek}, {Seager}, {Winn}, {Jenkins}, {Caldwell}, {Fong}, {Huang}, {Mireles}, {Schlieder}, {Shiao}, \& {Noel Villase{\~n}or}}]{Wong2021}
{Wong}, I., {Shporer}, A., {Zhou}, G., {et~al.} 2021, \aj, 162, 256

\bibitem[{{Yan} \& {Henning}(2018)}]{YanHenning2018}
{Yan}, F. \& {Henning}, T. 2018, Nature Astronomy, 2, 714

\bibitem[{{Yan} {et~al.}(2023){Yan}, {Nortmann}, {Reiners}, {Piskunov}, {Hatzes}, {Seemann}, {Shulyak}, {Lavail}, {Rains}, {Cont}, {Rengel}, {Lesjak}, {Nagel}, {Kochukhov}, {Czesla}, {Boldt-Christmas}, {Heiter}, {Smoker}, {Rodler}, {Bristow}, {Dorn}, {Jung}, {Marquart}, \& {Stempels}}]{Yan2023}
{Yan}, F., {Nortmann}, L., {Reiners}, A., {et~al.} 2023, \aap, 672, A107

\bibitem[{{Yan} {et~al.}(2022{\natexlab{a}}){Yan}, {Pall{\'e}}, {Reiners}, {Casasayas-Barris}, {Cont}, {Stangret}, {Nortmann}, {Molli{\`e}re}, {Henning}, {Chen}, \& {Molaverdikhani}}]{Yan2022b}
{Yan}, F., {Pall{\'e}}, E., {Reiners}, A., {et~al.} 2022{\natexlab{a}}, \aap, 661, L6

\bibitem[{{Yan} {et~al.}(2022{\natexlab{b}}){Yan}, {Reiners}, {Pall{\'e}}, {Shulyak}, {Stangret}, {Molaverdikhani}, {Nortmann}, {Molli{\`e}re}, {Henning}, {Casasayas-Barris}, {Cont}, {Chen}, {Czesla}, {S{\'a}nchez-L{\'o}pez}, {L{\'o}pez-Puertas}, {Ribas}, {Quirrenbach}, {Caballero}, {Amado}, {Galad{\'\i}-Enr{\'\i}quez}, {Khalafinejad}, {Lara}, {Montes}, {Morello}, {Nagel}, {Sedaghati}, {Zapatero Osorio}, \& {Zechmeister}}]{Yan2022a}
{Yan}, F., {Reiners}, A., {Pall{\'e}}, E., {et~al.} 2022{\natexlab{b}}, \aap, 659, A7

\bibitem[{{Yang} {et~al.}(2024){Yang}, {Chen}, {Yan}, {Tan}, \& {Ji}}]{Yang2024}
{Yang}, Y., {Chen}, G., {Yan}, F., {Tan}, X., \& {Ji}, J. 2024, \apjl, 971, L8

\bibitem[{{Zechmeister} {et~al.}(2014){Zechmeister}, {Anglada-Escud{\'e}}, \& {Reiners}}]{Zechmeister2014}
{Zechmeister}, M., {Anglada-Escud{\'e}}, G., \& {Reiners}, A. 2014, \aap, 561, A59

\bibitem[{{Zhang} {et~al.}(2022){Zhang}, {Snellen}, {Wyttenbach}, {Nielsen}, {Lendl}, {Casasayas-Barris}, {Chaverot}, {Kesseli}, {Lovis}, {Pepe}, {Psaridi}, {Seidel}, {Udry}, \& {Ulmer-Moll}}]{Zhang2022}
{Zhang}, Y., {Snellen}, I. A.~G., {Wyttenbach}, A., {et~al.} 2022, \aap, 666, A47

\end{thebibliography}


\appendix

\section{Posterior distributions}

\FloatBarrier

\begin{figure}[htbp]
    \centering
    \includegraphics[width=\textwidth]{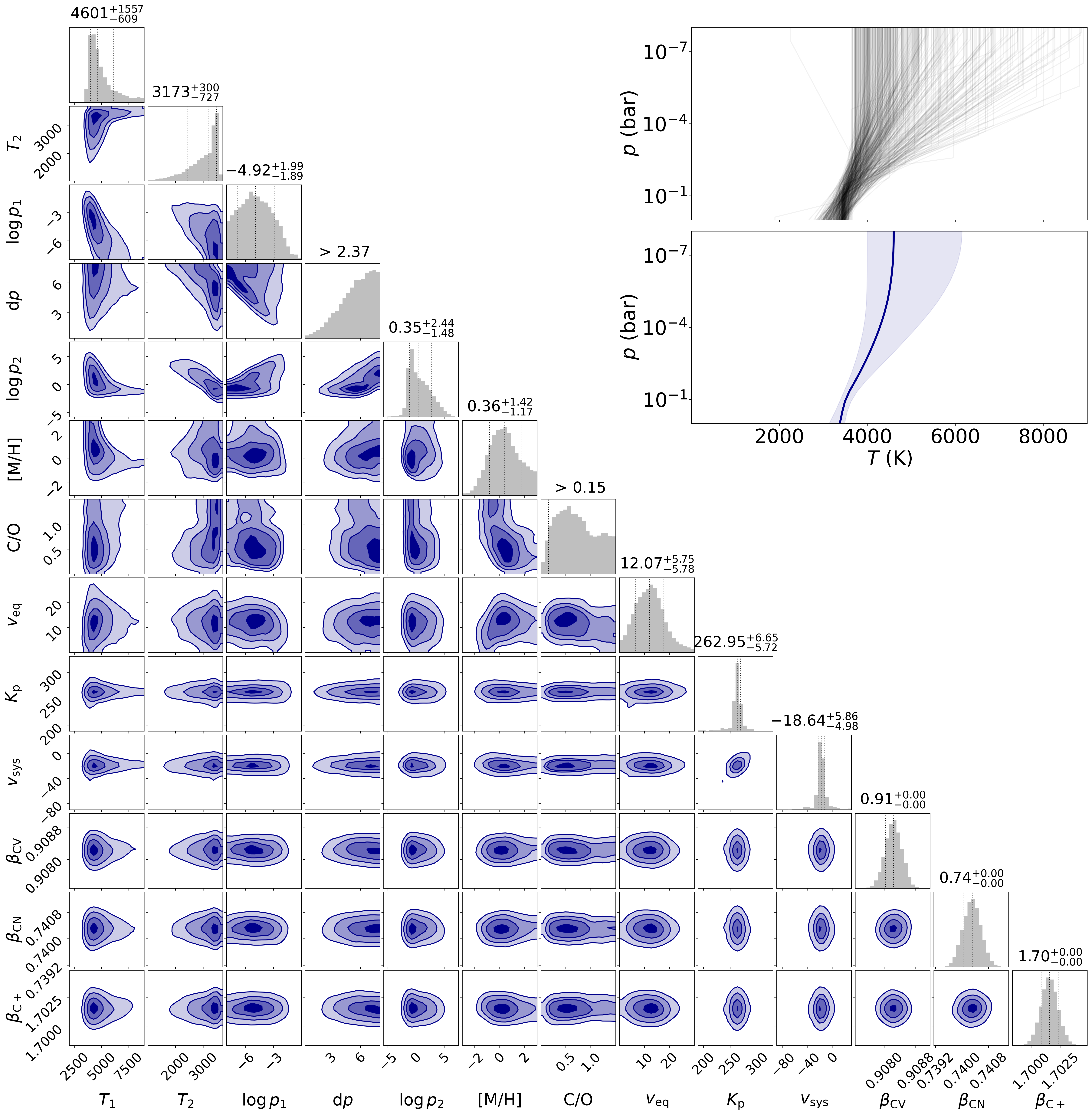}
    \begin{minipage}{\textwidth}
        \vspace{0.35cm}
        \captionof{figure}{Results of retrieval investigating the dayside atmosphere of TOI-2109b. {\it Corner plot}: posterior distributions and correlations between the atmospheric parameters. The dashed vertical lines in the posterior distributions denote the median and 1$\sigma$ intervals (16th, 50th, and 84th percentiles) for the bounded parameters. For parameters with lower limits, we report the 2$\sigma$ intervals (5th percentiles). We note that in addition to the free parameters of the retrieval, the present figure also includes $\log{p_2}$, which is calculated as the sum of the free parameters $\log{p_1}$ and $\mathrm{d}p$. {\it Top inset panel}: examples of the $T$-$p$ profiles sampled by the MCMC analysis. {\it Bottom inset panel}: median temperature curve with uncertainty intervals.}
        \label{figure:corner-plot-dayside}
    \end{minipage}
\end{figure}

\FloatBarrier

\begin{figure*}
        \centering
        \includegraphics[width=\textwidth]{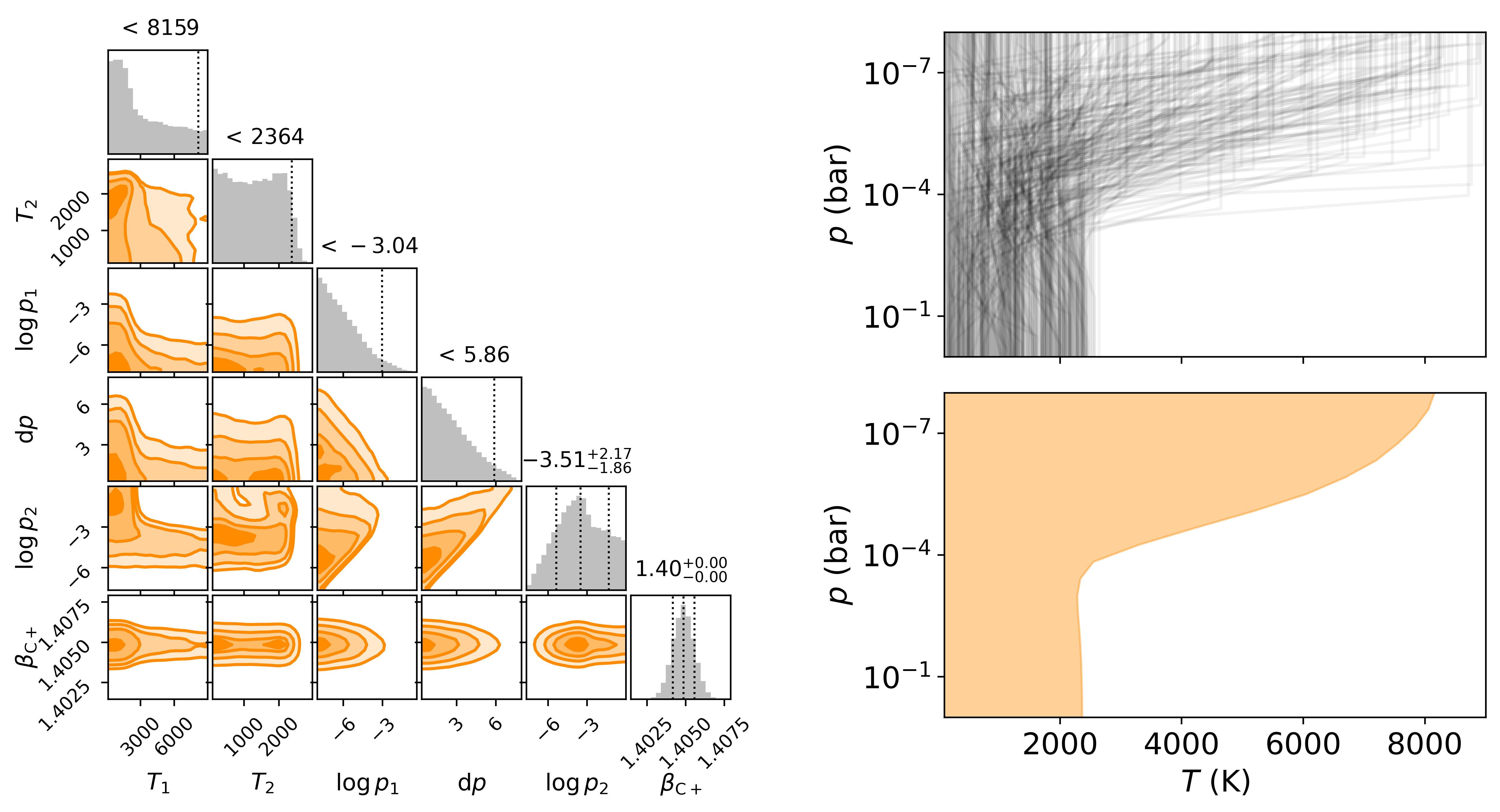}
        \caption{Results of retrieval investigating the nightside atmosphere of TOI-2109b. {\it Corner plot}: posterior distributions and correlations between the atmospheric temperature and uncertainty scaling parameters; chemical, velocity, and broadening parameters are set to fixed values. The dashed vertical lines in the posterior distributions denote the median and 1$\sigma$ intervals (16th, 50th and 84th percentiles) for the bounded parameters. For parameters with upper limits, we report the 2$\sigma$ intervals (95th percentiles). We note that in addition to the free parameters of the retrieval, the present figure also includes $\log{p_2}$, which is calculated as the sum of the free parameters $\log{p_1}$ and $\mathrm{d}p$. {\it Right top panel}: examples of the $T$-$p$ profiles sampled by the MCMC analysis. {\it Right bottom panel}: upper temperature limits of the nightside atmosphere as a function of atmospheric pressure.}
        \label{figure:corner-plot-nightside}
\end{figure*}

\end{document}